\documentclass{aa}
\usepackage{graphicx}
\usepackage[varg]{txfonts}
\usepackage{natbib}
\usepackage{soul}
\newlength{\linwx}
\usepackage{color}

\begin{document}

\title{Formation of planetary systems \\ by pebble accretion and migration: Growth of gas giants}
\author{
Bertram Bitsch \inst{1},
Andre Izidoro \inst{2},
Anders Johansen \inst{3},
Sean N. Raymond \inst{4},
Alessandro Morbidelli \inst{5},
Michiel Lambrechts \inst{3},
\and
Seth A. Jacobson \inst{6}
}
\offprints{B. Bitsch,\\ \email{bitsch@mpia.de}}
\institute{
Max-Planck-Institut f\"ur Astronomie, K\"onigstuhl 17, 69117 Heidelberg, Germany
\and
UNESP, Univ. Estadual Paulista - Grupo de Din\`amica Orbital Planetologia, Guaratinguet\`a, CEP 12.516-410, S\~ao Paulo, Brazil
\and
Lund Observatory, Department of Astronomy and Theoretical Physics, Lund University, 22100 Lund, Sweden
\and
Laboratoire d'Astrophysique de Bordeaux, CNRS and Universit\'e de Bordeaux, All\'ee Geoffroy St. Hilaire, 33165 Pessac, France
\and
University Nice-Sophia Antipolis, CNRS, Observatoire de la C\^{o}te d'Azur, Laboratoire LAGRANGE, CS 34229, 06304 Nice cedex 4, France
\and
Northwestern University, Department of Earth and Planetary Sciences, 2145 Sheridan Road, Evanston, IL 60208-3130, USA
}
\abstract{Giant planets migrate though the protoplanetary disc as they grow their solid core and attract their gaseous envelope. Previously, we have studied the growth and migration of an isolated planet in an evolving disc. Here, we generalise such models to include the mutual gravitational interaction between a high number of growing planetary bodies. We have investigated how the formation of planetary systems depends on the radial flux of pebbles through the protoplanetary disc and on the planet migration rate. Our N-body simulations confirm previous findings that Jupiter-like planets in orbits outside the water ice line originate from embryos starting out at 20-40 AU when using nominal type-I and type-II migration rates and a pebble flux of approximately 100-200 Earth masses per million years, enough to grow Jupiter within the lifetime of the solar nebula. The planetary embryos placed up to 30 AU migrate into the inner system ($r_{\rm P}<1$AU). There they form super-Earths or hot and warm gas giants, producing systems that are inconsistent with the configuration of the solar system, but consistent with some exoplanetary systems. We also explored slower migration rates which allow the formation of gas giants from embryos originating from the 5-10 AU region, which are stranded exterior to 1 AU at the end of the gas-disc phase. These giant planets can also form in discs with lower pebbles fluxes (50-100 Earth masses per Myr). We identify a pebble flux threshold below which migration dominates and moves the planetary core to the inner disc, where the pebble isolation mass is too low for the planet to accrete gas efficiently. In our model, giant planet growth requires a sufficiently-high pebble flux to enable growth to out-compete migration. An even higher pebble flux produces systems with multiple gas giants. We show that planetary embryos starting interior to 5 AU do not grow into gas giants, even if migration is slow and the pebble flux is large. These embryos instead grow to just a few Earth masses, the mass regime of super-Earths. This stunted growth is caused by the low pebble isolation mass in the inner disc and is therefore independent of the pebble flux. Additionally we show that the long term evolution of our formed planetary systems can naturally produce systems with inner super-Earths and outer gas giants as well as systems of giant planets on very eccentric orbits.
}
\authorrunning{Bitsch et al.}\maketitle

\section{Introduction}
\label{sec:Introduction}

The discovery of the first planet around another star yielded a surprise, because the detected planet was nothing like the planets in our own solar system \citep{1995Natur.378..355M}. The planet is in the mass regime of Jupiter and orbits its host star on a three-day orbit, which ultimately gave the name of this planetary class: hot Jupiters. Today we know that $\sim$1\% of solar like stars host hot Jupiter planets, while their cold analogues ($r_{\rm P}> 1$AU) are found around 10\% of stars \citep{J2010}. The occurrence rate of Jupiter planets in general seems to increase with their host star metallicity \citep{2005ApJ...622.1102F}. Additionally, the eccentricity distribution of these giant planets also increases with host star metallicity, meaning that giant planets are more likely on eccentric orbits if the metallicity is large \citep{2018arXiv180206794B}, an attribute associated with the formation of multiple giant planets. Nevertheless, the exact growth mechanism of giant planet systems still remains a mystery.

In classical simulations of the core accretion scenario, km-sized planetesimals can grow through mutual collisions to form the cores of giant planets \citep{1996Icar..124...62P}. However, to achieve core masses of a few Earth masses, planetesimal densities of multiple times the solid density invoked in the minimum mass solar nebular (MMSN, \citealt{1977Ap&SS..51..153W, 1981PThPS..70...35H}) are needed, to achieve growth timescales compatible with the disc lifetime. In addition, the growth time-scale increases with orbital distance, making the formation of planetary cores at large orbital distances very hard \citep{2003Icar..161..431T}. Gravitational stirring of the planetesimals by a set of growing protoplanets reduces the growth rates even further \citep{2010AJ....139.1297L}. Also protoplanets migrating through a sea of planetesimals and planetary embryos mostly scatter these bodies and accretion is inefficient \citep{1999Icar..139..350T, 2007ApJ...660..823M, 2014ApJ...794...11I}.

However, by taking the accretion of small mm-cm sized pebbles into account, the growth time-scale of planetary cores can be greatly reduced \citep{2010MNRAS.404..475J, 2010A&A...520A..43O, 2012A&A...544A..32L, 2012A&A...546A..18M}. The efficiency of pebble accretion relies on the drift of the pebbles in the gas discs. When a pebble enters within the planetary Bondi or Hill sphere, the gas drag robs the pebble of angular momentum and allows it to spiral inward onto the planet to be accreted. This effect becomes important once the planetesimal has reached several 100 km in size \citep{2016A&A...591A..72I, 2016A&A...586A..66V, Johansen2017}.

Growing planets gravitationally interact with the protoplanetary disc and migrate through it \citep{1986Icar...67..164W, 1997Icar..126..261W}, where the planet only slightly influences the gas distribution around it. Low mass bodies migrate in type-I migration, which is mostly inwards \citep{2006A&A...459L..17P, 2009A&A...506..971K}. The migration time-scale related to type-I migration can be orders of magnitude shorter than the disc lifetime, driving the growing planets to the central star \citep{2002ApJ...565.1257T}. Jupiter-mass planets, on the other hand, open deep gaps in the protoplanetary disc \citep{1986ApJ...309..846L, 2006Icar..181..587C}. The planet then migrates in type-II migration, which is slower than type-I-migration and has the potential of saving the growing planets from migrating all the way to the central star. New studies also indicate that the classical type-II migration operating on a viscous time-scale, might not be valid \citep{2014ApJ...792L..10D, 2015A&A...574A..52D, 2018arXiv180511101K, 2018arXiv180800381R}. We will thus investigate here the influences of these new type-II migration prescriptions on planet formation.

Pebble accretion in combination with disc evolution and planet migration is used to study the formation of planetary systems in simulations with single bodies \citep{2015A&A...582A.112B, 2016A&A...590A.101B, 2017ASSL..445..339B, 2018A&A...609C...2B, 2018MNRAS.474..886N, 2018arXiv181100523J}. These results indicate that pebble accretion is efficient enough for planets to reach pebble isolation mass before migrating due to type-I migration all the way to the central star. At pebble isolation mass, the planet opens a partial gap in the disc, increasing the gas velocity exterior of its orbit to super-Keplerian values that prevent the inward drift of pebbles onto the planet, which thus stops accreting pebbles \citep{2006A&A...453.1129P, 2012A&A...546A..18M, 2014A&A...572A..35L, 2018arXiv180102341B} and gas accretion can start. However, cold Jupiters (orbits with $a>1$ AU) migrate over large orbital distances during their formation \citep{2015A&A...582A.112B, 2016A&A...590A.101B, 2017ASSL..445..339B, 2018MNRAS.474..886N}.

Additionally, N-body studies including pebble accretion are employed to explain the evolution of multiple embryos into planetary systems \citep{2015Natur.524..322L, 2016ApJ...825...63C, 2017A&A...607A..67M, 2018arXiv180405510M}. \citet{2015Natur.524..322L} found that only a few bodies will grow to become giant planets, because the largest planetary embryos excite the eccentricities and inclinations of their smaller counterparts to such high values that their ability to accrete pebbles is quenched. Similar results were obtained also by \citet{2018arXiv180405510M}, who along with \citet{2015Natur.524..322L} ignore planet migration. The simulations by \citet{2017A&A...607A..67M}, on the other hand, showed that embryos starting at the current orbital distances of Jupiter and Saturn migrate to the inner disc while they grow. This is not surprising, since \citet{2015A&A...582A.112B} and \citet{2018A&A...609C...2B} showed that the formation of gas giants at a few AU distance requires the seed to start growing beyond 20-40 AU in the disc due to the short migration time-scales.

In this work we have investigated the conditions that allow the formation of cold gas giants outside of 1 AU by investigating two main parameters that determine the outcome of the simulations: (i) the pebble flux that determines the growth rate of the planetary core and (ii) the migration speed during planet growth which determines the final semi-major axis of the planet. 

This paper is part of a trilogy. The other two companion papers focus on the difference and similarities between truly Earth-like planets and rocky super-Earths \citep{Lambrechts18} as well as on the formation of super-Earth systems \citep{Izidoro18} via pebble accretion, planet migration and breaking of resonance chains. In those two works, the pebble flux plays a crucial role as it determines the divide between the formation of true terrestrial planet analogues and super-Earths up to planets in the ice giant mass regime (15-20 Earth masses).

Our work is structured as follows. In Section~\ref{sec:method} we describe our new code FLINTSTONE, which includes pebble and gas accretion as well as planet migration in a protoplanetary disc environment, which is also used in \citet{Izidoro18}. We particularly focus here on the gas accretion recipes and the migration prescriptions. In Section~\ref{sec:nominal} we discuss the influence of the pebble flux as well as the importance of the initial position of the planetary embryo on the formation of gas giants and verify previous results of single-body simulations. In Section~\ref{sec:Kanagawamig} we show that an early gap opening and a reduced migration speed in the type-II migration regime allows the formation of giant planet systems, originating much closer to the central star, if the pebble flux is large enough. We then show outcomes of individual systems and long term evolution in Section~\ref{sec:systems} and discuss our findings in Section~\ref{sec:disc}. We finally summarise in Section~\ref{sec:summary}.

\section{Method}
\label{sec:method}

We used our new pebble accretion and N-body code FLINTSTONE, which includes prescriptions for planet migration, gas damping of eccentricity and inclination and disc evolution. The original N-body integrator is based on the Mercury hybrid symplectic integrator \citep{1999MNRAS.304..793C} and collisions between embryos were treated as inelastic mergers. Migration and type-I damping of eccentricities and inclinations \citep{2000MNRAS.315..823P, 2004ApJ...602..388T} was included using the equations of \citet{2008A&A...482..677C}. For massive planets we use the damping rates derived by \citet{2013A&A...555A.124B}. Except for the pebble accretion, gas accretion and new migration schemes (discussed below), this code was already used by \citet{2017MNRAS.470.1750I} and \citet{Raymond2018} to explain the formation and evolution of super-Earth systems. Additionally the same code is used in the companion paper of \citet{Izidoro18}, which is described there in more detail. We thus give here just a quick summary of the methods used and highlight the differences to our companion papers.

We additionally performed test simulations of single bodies regarding planet growth via pebble accretion and planet migration and found perfect match with the N-body code used in \citet{Lambrechts18}, which is based on SyMBA \citep{1998AJ....116.2067D}.

\subsection{Set-up of the models}

We initialise our planetary seeds in the mass range of 0.005-0.015 Earth masses, which is roughly the planetary mass at which pebble accretion becomes more efficient than planetesimal accretion for single bodies \citep{Johansen2017}. Our planetary cores have a fixed density of 2.0 g/cm$^3$, corresponding to a mixture of water ice and rock. Our initial seeds are spaced in two configurations: (i) where the distance between the seeds is fixed at 0.25 AU and the innermost seed is placed at 2.75 AU to study the formation of planets in the inner disc and (ii) a distance of 10-15 mutual Hill radii where the innermost seed is placed at 10 AU to study the formation of planets in the outer system. In both configurations, the initial eccentricity of the planetary embryo is 0.001-0.01, and the inclination is 0.01-0.5 degrees and with random orbital elements.

Our disc model and evolution is based on the simulations by \citet{2015A&A...575A..28B}, where we implant our planetary embryos in a disc that is already 2 Myr old (in the model of \citet{2015A&A...575A..28B}) and lives for a total of 5 Myr. The planets thus evolve for 3 Myr in the gas disc environment. In our disc model, this corresponds to an accretion rate of $5\times10^{-9}{\rm M}_\odot/{\rm yr}$ at the beginning of the simulation and of $1\times10^{-9}{\rm M}_\odot/{\rm yr}$ at disc dissipation. A disc with a low accretion rate has a low gas surface density resulting in slow type-I migration, which we identify below as a main problem for gas giant formation at large distances. We note that our simulations are mostly independent of the chosen disc accretion rate, as long as the pebble isolation mass is larger in the outer disc than in the inner disc, which is true for early discs older than 0.5 Myr \citep{2015A&A...575A..28B}. In this scenario, the planets in the outer disc can grow to gas giants, while the inner planets can only grow to super-Earths. What then matters for the final system is the pebble flux and migration speed, which determine the exact configuration of the formed systems\footnote{ As shown in our companion paper by \citet{Izidoro18}, hot inner discs are required to make rocky super-Earths, but this has only very little influence on the formation of outer gas giants that accrete mainly icy pebbles.}. The here used discs with low accretion rates are on the lower end of the observed accretion rates, which spread over a large range \citep{2016A&A...585A.136M}. See also section~\ref{sec:disc} for more details.

Discs with these low accretion rates already feature very small aspect ratios in the inner few AU, resulting in very low pebble isolation masses and very low core masses (see appendix~\ref{ap:Miso}). Additionally, the water ice line is at $\sim$1 AU at the beginning of our simulations and moves down to $\sim 0.5$ AU at the end of the gas disc phase. This implies that we focus on the growth of giant planets in the cold parts of the disc, where the pebbles are larger due the water ice component (see below).

The viscosity $\nu$ in our work is described through the $\alpha$-model \citep{1973A&A....24..337S}. The viscosity in itself is a crucial parameter, because it determines the disc structure, the pebble accretion rates (through the pebble scale height $H_{\rm peb}$) and the planet migration rates. The disc structure of \citet{2015A&A...575A..28B} is calculated using $\alpha_{\rm disc}=0.0054$, which is fixed to that value in the \citet{2015A&A...575A..28B} thermodynamical model and is kept constant in our work her for all shown simulations.

In our work we test different viscosities related to migration, $\nu_{\rm mig}$, while keeping the viscosity for the disc parameters and pebble growth, $\nu_{\rm disc}$, constant. Although this is not self-consistent, it has the advantage of letting us probe just the influence of migration without changing the disc structure and planetary accretion rates. Due to the $\alpha$-parameterization we link the disc's viscosity to $\alpha_{\rm disc}$ and the viscosity for migration to $\alpha_{\rm mig}$ and thus only vary $\alpha_{\rm mig}$ in this work.

We believe that this approach has merit, because new disc models are emerging, where the radial transport of gas is dominated by the angular momentum removal in MHD winds at the surface of the disc. These discs may have a structure in terms of surface density profile similar to that of $\alpha$-discs with relatively large viscosity \citep{2017arXiv170700729B}, though having a low bulk viscosity. Giant planets in these discs may migrate at low velocity \citep{2018ApJ...864...77I, 2018arXiv181100523J}, which we can mimic here by reducing the migration viscosity parameter $\alpha_{\rm mig}$. This can potentially allow us to obtain results consistent with observations (i.e. that most giant planets are 'cold').

\subsection{Pebble accretion}
\label{subsec:accretion}

We follow the prescription of pebble accretion from \citet{Johansen2015}, which includes a decrease in the pebble accretion for objects on eccentric or inclined orbits. The accretion rate of the planetary core $\dot{M}_{\rm core}$ is directly proportional to the pebble surface density $\Sigma_{\rm peb}$. 

We would like to stress that in the companion paper by \citet{Lambrechts18} the pebble flux $\dot{M}_{\rm peb}$ is treated as a free parameter, decaying as the gas disc. This simpler prescription is suited for the scopes of \citet{Lambrechts18} defining the transition between terrestrial planets and super-Earths, while we use here and in \citet{Izidoro18} a more quantitative model for the formation of planetary systems, based on \citet{2014A&A...572A.107L}, which was also used in \citet{2015A&A...582A.112B} and \citet{2018A&A...609C...2B}. As the details of our pebble accretion scheme are also described in our companion paper by \citet{Izidoro18}, we will only mention the necessary and important parameters for this work. 

Pebbles in the protoplanetary disc settle towards the midplane depending on their size, parameterised in this work by the dimensionless Stokes number $\tau_{\rm f}$, depending on the level of turbulence in the protoplanetary disc described through the disc's viscosity. Using the $\alpha$ prescription, \citet{2007Icar..192..588Y} derived the pebble scale height as
\begin{equation}
 H_{\rm peb} = H_{\rm g} \sqrt{\alpha_{\rm disc} / \tau_{\rm f}} \ .
\end{equation}
Typically the pebbles in our simulations have Stokes numbers of 0.05-0.1, which is calculated by equating the drift time-scale with the growth time-scale \citep{2012A&A...539A.148B, 2012A&A...544A..32L}. That yields a value of $H_{\rm peb}/H_{\rm g}\sim$0.1, in agreement with observations of protoplanetary discs \citep{2016ApJ...816...25P}.

\begin{figure}
 \centering
 \includegraphics[scale=0.7]{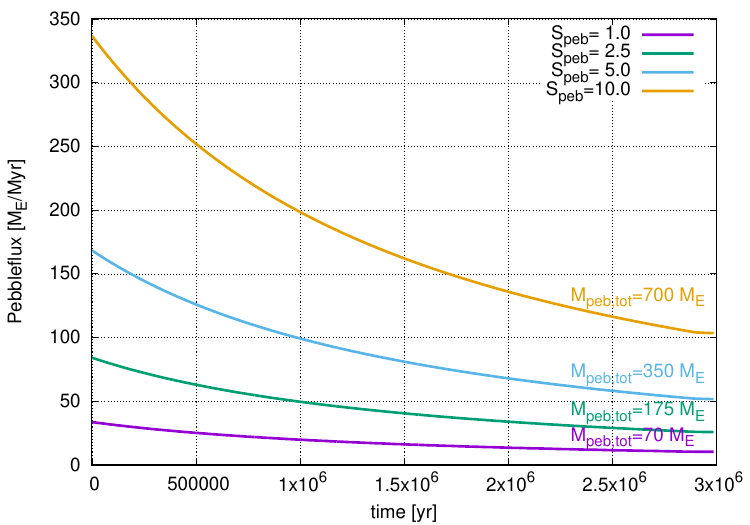}
 \caption{Pebble flux as a function of time in our disc model with a gas disc lifetime of 3 Myr. We also show the upscaled pebble fluxes using the factor $S_{\rm peb}=1.0-10.0$. Additionally, the total amount of pebbles $M_{\rm peb,tot}$ drifting through the gas disc during its lifetime of 3 Myr is marked.
   \label{fig:pebbleflux}
   }
\end{figure}

The pebble surface density $\Sigma_{\rm peb} (r_{\rm P})$ at the planets location can be calculated from the pebble flux $\dot{M}_{\rm peb}$ via
\begin{equation}
 \label{eq:SigmaPeb}
 \Sigma_{\rm peb} (r_{\rm P}) = \sqrt{\frac{2 S_{\rm peb} \dot{M}_{\rm peb} \Sigma_{\rm g}(r_{\rm P}) }{\sqrt{3} \pi \epsilon_{\rm P} r_{\rm P} v_{\rm K}}} \ ,
\end{equation}
where $r_{\rm P}$ denotes the semi-major axis of the planet, $v_{\rm K} = \Omega_{\rm K} r_{\rm P}$, and $\Sigma_{\rm g} (r_{\rm P})$ stands for the gas surface density at the planets locations. The pebble flux $\dot{M}_{\rm peb}$ is calculated self consistently through an equilibrium between dust growth and drift \citep{2012A&A...539A.148B, 2014A&A...572A.107L, 2018A&A...609C...2B}, where these simulations predict a decrease of the pebble flux in time (Fig.~\ref{fig:pebbleflux}). Here $S_{\rm peb}$ describes the scaling factor to the pebble flux $\dot{M}_{\rm peb}$ to test the influence of different pebble fluxes (see below). The same approach is used in \citet{Izidoro18}. The pebble sticking efficiency can be taken as $\epsilon_{\rm P} =0.5$ under the assumption of near-perfect sticking \citep{2014A&A...572A.107L}. 

The Stokes number of the pebbles can be related to the pebble surface density $\Sigma_{\rm peb}$ and gas surface density $\Sigma_{\rm g}$ through the following relation
\begin{equation}
 \tau_{\rm f} = \frac{\sqrt{3}}{8} \frac{\epsilon_{\rm P}}{\eta} \frac{\Sigma_{\rm peb}(r_{\rm P})}{\Sigma_{\rm g} (r_{\rm P})} \ .
\end{equation}
Here $\eta$ represents a measurement of the sub-Keplerianity of the gas velocity.

We note that we did not include here the effects of an outer reservoir of pebbles\footnote{ An outer reservoir of pebbles or a constant replenishment of pebbles might be needed to keep the pebble flux large enough to allow the formation of planets \citep{2018arXiv180907374M}.} in the disc as proposed in \citet{2018A&A...609C...2B}, but follow the reduction of the pebble flux in time as predicted directly by the pebble evolution models (see Fig.~\ref{fig:pebbleflux}). This yields $\dot{M}_{\rm peb}$ and thus $\Sigma_{\rm peb}$ of values lower than predicted by observations \citep{2018A&A...609C...2B}. These low $\Sigma_{\rm peb}$ values result in very low pebble accretion rates, making the formation of giant planet cores difficult. We thus vary the pebble flux by a factor $S_{\rm peb}$ with 2.5, 5.0, 10.0 in order to achieve higher accretion rates, where we show the evolution of $\dot{M}_{\rm peb}$ in Fig.~\ref{fig:pebbleflux}. $S_{\rm peb}=1.0$ thus corresponds to the nominal pebble flux. Larger $S_{\rm peb}$ values thus lead to pebble surface densities that correspond to the observed values \citep{2011ARA&A..49...67W, 2016ApJ...821L..16C}. Using $S_{\rm peb}=10.0$ results in Stokes numbers that are a factor of $\sim$3 larger than for our nominal pebble flux, resulting in faster accretion rates due to the larger pebbles and larger $\Sigma_{\rm peb}$. The same approach has been taken in our companion paper by \citet{Izidoro18}.

At higher temperatures, water sublimates and we fix the radius of the pebbles to 1 mm, corresponding to typical chondrule sizes \citep{2015Icar..258..418M, 2016A&A...591A..72I}. This is consistent with the assumptions made in \citet{2015Icar..258..418M} to explain the dichotomy between the terrestrial planets and the gas giants in the solar system. Fixing the pebble size to 1 mm, however still yields Stokes numbers of various values, due to the radially declining gas surface density. Additionally, we reduce the pebble flux $\dot{M}_{\rm peb}$ to half its nominal value to account for water loss. In our disc model, the water ice line is located at $\approx$1 AU at the beginning of our simulations, but moves even further inwards in time as the disc evolves \citep{2015A&A...575A..28B}.

In our companion paper by \citet{Lambrechts18} a total pebble mass of $\approx 170 {\rm M}_{\rm E}$ over the disc lifetime is needed to form super-Earth type planets by the accretion of pebbles, while lower pebble fluxes in their work do not result in sufficient growth to form super-Earths directly from pebbles. This corresponds to $S_{\rm peb} = 5.0$ in our model, because of the volatile loss at the water ice line when silicate pebbles drift into the inner system as required by \citet{Lambrechts18}. This pebble flux is sufficient in our case for gas giant growth. The reason for this difference is that \citet{Lambrechts18} investigate the formation of rocky super-Earths that form interior to the water ice line, where the pebbles are small. This results then in lower accretion rates compared to the pebble sizes used in our work here, which are about an order of magnitude larger than in \citet{Lambrechts18}.

A planet accretes a fraction $f_{\rm acc}$ of the whole pebble flux $\dot{M}_{\rm peb}$ passing it
\begin{equation}
\label{eq:facc}
 f_{\rm acc} = \frac{\dot{M}_{\rm core}}{\dot{M}_{\rm peb}} \ .
\end{equation}
The pebble flux arriving at interior planets is thus reduced by exactly this fraction $f_{\rm acc}$, reducing the accretion rates onto the interior planets. For low core masses $f_{\rm acc}$ is very small and stays well below a level of a percent. Even larger cores do not reduce the pebble flux significantly, until they reach pebble isolation mass \citep{2012A&A...546A..18M, 2014A&A...572A..35L, 2018arXiv180102341B}, when pebble accretion stops (see below) and the whole inward flux of pebbles is stopped, starving the inner planetary embryos.

As a planet grows, it starts to push away material from its orbit, generating a partial gap in the protoplanetary disc, where the planet generates an inversion in the radial pressure gradient of the disc exterior its orbit, halting the inward drift of pebbles \citep{2006A&A...453.1129P, 2012A&A...546A..18M, 2014A&A...572A..35L, 2018arXiv180102341B, 2018A&A...615A.110A}. The pebble isolation mass in itself is a function of the local properties of the protoplanetary discs, namely the disc's viscosity $\nu$, aspect ratio $H/r$ and radial pressure gradient $\partial \ln P/\partial\ln r$ as well as of the Stokes number of the particles, which can diffuse through the partial gap in the disc generated by the planet \citep{2018arXiv180102341B}. We follow here the fit of \citet{2018arXiv180102341B}, who gave the pebble isolation mass including diffusion of small pebbles as
\begin{equation}
\label{eq:MisowD}
  M_{\rm iso} = 25 f_{\rm fit} {\rm M}_{\rm E} + \frac{\Pi_{\rm crit}}{\lambda} {\rm M}_{\rm E} \ ,
\end{equation}
with $\lambda \approx 0.00476 / f_{\rm fit}$, $\Pi_{\rm crit} = \frac{\alpha_{\rm disc}}{2\tau_{\rm f}}$, and
\begin{equation}
\label{eq:ffit}
 f_{\rm fit} = \left[\frac{H/r}{0.05}\right]^3 \left[ 0.34 \left(\frac{\log(\alpha_3)}{\log(\alpha_{\rm disc})}\right)^4 + 0.66 \right] \left[1-\frac{\frac{\partial\ln P}{\partial\ln r } +2.5}{6} \right] \ ,
\end{equation}
where $\alpha_3 = 0.001$. In appendix~\ref{ap:Miso} we show how the pebble isolation mass evolves in our disc model in time.

\subsection{Gas accretion}

After the planets have reached pebble isolation mass, the planetary envelope can contract, where we follow the analytical prescriptions of \citet{2014ApJ...786...21P} and \citet{2015ApJ...800...82P}. In this formalism, the contraction rate of the planetary envelope is a strong function of the planetary core mass
\begin{eqnarray}
\label{eq:Mdotenv}
 \dot{M}_{\rm gas} &= 0.000175 f^{-2} \left(\frac{\kappa_{\rm env}}{1{\rm cm}^2/{\rm g}}\right)^{-1} \left( \frac{\rho_{\rm c}}{5.5 {\rm g}/{\rm cm}^3} \right)^{-1/6} \left( \frac{M_{\rm c}}{{\rm M}_{\rm E}} \right)^{11/3} \nonumber \\ 
 &\left(\frac{M_{\rm env}}{{\rm M}_{\rm E}}\right)^{-1} \left( \frac{T}{81 {\rm K}} \right)^{-0.5} \frac{{\rm M}_{\rm E}}{{\rm Myr}} \ .
\end{eqnarray}
Here $f$ is a factor to change the accretion rate in order to match numerical and analytical results, which is normally set to $f=0.2$ \citep{2014ApJ...786...21P}. The opacity in the planetary envelope $\kappa_{\rm env}$ is generally very hard to determine because it depends on the grain sizes, their composition and their distribution inside the planetary atmosphere \citep{2014A&A...572A.118M}. Here we use $\kappa_{\rm env} = 0.05{\rm cm}^2/{\rm g}$, which is very similar to the values used in the study by \citet{2008Icar..194..368M}. Lower values of $\kappa_{\rm env}$ yield faster contraction rates, while higher $\kappa_{\rm env}$ result in slower envelope contraction. This contraction phase ends as soon as $M_{\rm core}=M_{\rm env}$ and rapid gas accretion starts.

For rapid gas accretion ($M_{\rm core} < M_{\rm env}$), we follow \citet{2010MNRAS.405.1227M}, who calculated the gas accretion rate using 3D hydrodynamical simulations with nested grids. They find two different gas accretion branches, which are given as 
\begin{equation}
 \dot{M}_{\rm gas,low} = 0.83 \Omega_{\rm K} \Sigma_{\rm g} H^2 \left( \frac{r_{\rm H}}{H} \right)^{9/2}
\end{equation}
and
\begin{equation}
 \dot{M}_{\rm gas,high} = 0.14 \Omega_{\rm K} \Sigma_{\rm g} H^2 \ ,
\end{equation}
where $r_{\rm H}$ denotes the planetary hill radius. The effective accretion rate is given by the minimum of these two accretion rates. The low branch is for low mass planets (with $r_{\rm H}/H < 0.3$), while the high branch is for high mass planets ($r_{\rm H}/H > 0.3$). Additionally, we limit the maximum accretion rate to $80\%$ of the disc's accretion rate onto the star, because gas can flow through the gap, even for high mass planets \citep{2006ApJ...641..526L}. The final masses of the planets is limited to Jupiters mass, as we are mainly interested in the formation of the giant planets in our solar system\footnote{In the simulations using the minimal pebble flux to allow gas giant formation in our model, $S_{\rm peb}=2.5$, planets barely reach Jupiter mass at the end of the gas discs lifetime, so we think that this limitation does not influence our results.}.

Previous hydrodynamical simulations using the \citet{2010MNRAS.405.1227M} gas accretion rates have shown that planets first accrete the material directly from their horseshoe region \citep{2017Icar..285..145C}. Only after they accreted all the gas contained in that region is accretion limited by what the disc can provide. In our simulations, however, we limit gas accretion always to 80\% of the disc's accretion rate $\dot{M}_{\rm disc}$. Additionally multiple gas accreting planets would in reality compete for the gas of the disc, but in our model planets in the runaway gas accretion mode accrete gas to the full rate, limited to 80\% of the disc's accretion rate. For simplification we also do not include the feedback of a gas accreting planet onto the disc structure, but we will include this in future work.

\subsection{Nominal type-I and type-II migration}
\label{subsec:nominal}

Planets growing in protoplanetary discs interact gravitationally with the disc and migrate through it (see \citealt{2013arXiv1312.4293B} for a review). The exact prescriptions of type-I migration and type-I damping of eccentricity and inclination are described in our companion paper of \citet{Izidoro18}. We will thus just repeat the necessities and differences between our work and \citet{Izidoro18}

The entropy related corotation torque can drive outward migration in the inner regions of the disc \citep{2015A&A...575A..28B}, if the viscosity is high enough. This is for example the case for $\alpha_{\rm mig}=0.0054$ as used in our nominal migration model and in our companion paper by \citet{Izidoro18}. However for low levels of viscosity the entropy related corotation torque saturates and it becomes negative, preventing outward migration. This is the case in our simulations with low viscosity, using $\alpha_{\rm mig}=0.00054$ and $\alpha_{\rm mig}=0.0001$.

In \citet{Izidoro18} planets do not accrete gaseous envelopes and thus stay relatively small and always migrate in the type-I regime. Here we include gas accretion and thus also include type-II migration of giant planets. The classical type-II migration depends solely on the disc's viscosity \citep{1986ApJ...309..846L}, even though there has been some debate about this in recent literature \citep{2015A&A...574A..52D, 2017A&A...598A..80D, 2018MNRAS.474.4460R, 2018arXiv180800381R}. Type-II migration sets in when the planet has opened a deep gap in the protoplanetary disc. For a planet to open a gap, the gap opening criteria by \citet{2006Icar..181..587C} has to be fulfilled, which is given as
\begin{equation}
\label{eq:gapopen}
 \mathcal{P} = \frac{3}{4} \frac{H}{r_{\rm H}} + \frac{50}{q \mathcal{R}} \leq 1 \ .
\end{equation}
Here $q$ is the star to planet mass ratio, and $\mathcal{R}$ the Reynolds number given by $\mathcal{R} = r_{\rm P}^2 \Omega_{\rm P} / \nu$. Fulfilling this relation leads to surface density at the bottom of the gap $\Sigma_{\rm min}$ that corresponds to 10\% of the unperturbed gas surface density $\Sigma_{\rm up}$. The migration time-scale is then given by $\tau_{\rm visc} = r_{\rm P}^2 / \nu$. As in \citet{2015A&A...582A.112B} we use an interpolation between type-I and type-II migration in order to smooth the transition.

As we are mostly interested in the growth of giant planets exterior to 1 AU, we remove inward migrating bodies that cross interior to 1 AU from our simulations, in order to save computation time. The removed bodies should also have an eccentricity lower than $e<0.5$, in order to avoid removing bodies that have large eccentricity due to a scattering event and just cross inside this boundary temporarily. In reality these bodies would survive in the inner disc, where they can pile-up close to the star and form systems of hot super-Earths, like in our companion paper by \citet{Izidoro18} and in section~\ref{sec:systems}.

\subsection{Migration with reduced viscosity}
\label{subsec:kana}

Recent hydrodynamical simulations have cast doubt on the hypothesis that type-II migration follows the viscous evolution of the protoplanetary disc \citep{2014ApJ...792L..10D, 2015A&A...574A..52D, 2017A&A...598A..80D, 2018arXiv180511101K, 2018arXiv180800381R}. We first study the evolution of systems with the nominal migration rates (see above) in section~\ref{sec:nominal} that correspond to the migration rates of \citet{2015A&A...582A.112B}. We then show the influence of a reduced $\alpha_{\rm mig}$ parameter on the formation of planetary systems in section~\ref{sec:Kanagawamig}, where the differences in migration speed are described below. A similar approach also following the work of \citet{2018arXiv180511101K}, as outlined below, has been used in the planet population synthesis models by \citet{2018ApJ...864...77I}. With the reduced migration rates they find a good match to the cold Jupiter distribution as revealed by observations.

Recent 2D simulations by \citet{2018arXiv180511101K} showed that the torque exerted on a gap-opening planet depends on the surface density at the bottom of the gap, $\Sigma_{\rm min}$. The planet slows down as the surface density at the bottom of the gap decreases. \citet{2018arXiv180511101K} provide a migration formula for gap-opening planets that allow slow migration, if the viscosity is low. We will model this reduced migration rates using the before introduced $\alpha_{\rm mig}$ with different values.

\citet{2018arXiv180511101K} relate the type-II migration rate to the type-I migration time-scale (which we calculate as explained above) in the following way
\begin{equation}
\label{eq:migII}
 \tau_{\rm mig II} = \frac{\Sigma_{\rm up}}{\Sigma_{\rm min}} \tau_{\rm mig I} \ ,
\end{equation}
where $\Sigma_{\rm up}$ corresponds to the unperturbed gas surface density and $\Sigma_{\rm min}$ to the minimal gas surface density at the bottom of the gap generated by the planet. The ratio $\Sigma_{\rm up}/\Sigma_{\rm min}$ can be expressed through \citep{2013ApJ...769...41D, 2014ApJ...782...88F, 2015MNRAS.448..994K}
\begin{equation}
\label{eq:Kgapopen}
 \frac{\Sigma_{\rm up}}{\Sigma_{\rm min}} = 1 + 0.04 K \ ,
\end{equation}
where
\begin{equation}
 K = \left( \frac{M_{\rm P}}{{\rm M}_\odot} \right)^2 \left( \frac{H}{r} \right)^{-5} \alpha_{\rm mig}^{-1} \ .
\end{equation}
This prescription naturally allows a smooth transition from type-I to type-II migration as the planet grows, while the transition between type-I and type-II migration used in \citet{2015A&A...582A.112B} originates from connecting different sets of simulations. Additionally this prescription allows also an earlier transition into type-II migration due to the lower viscosity, which reduces the distance planets migrate. This effect can not be modelled by just reducing type-II migration alone, because gap opening would occur only for planets with very high masses, if $\alpha$ in eq.~\ref{eq:gapopen} is large. We show this behaviour in appendix~\ref{ap:migration}.

In the following we use two different migration prescriptions:
\begin{enumerate}
 \item The nominal migration prescription already used in \citet{2015A&A...582A.112B}. This corresponds to the simulations shown in section~\ref{sec:nominal}.
 \item The migration prescription using eq.~\ref{eq:migII} for type-II migration and the transition from type-I to type-II migration using $\alpha_{\rm disc}=0.00054$ or $\alpha_{\rm disc}=0.0001$ in the other sections.
\end{enumerate}
We note that using $\alpha_{\rm mig} = \alpha_{\rm disc} = 0.0054$ with eq.~\ref{eq:migII} for migration results in a nearly identical migration history as for simulations of planets using the nominal migration prescription (used in section~\ref{sec:nominal}) and classical type-II migration rates used in \citet{2015A&A...582A.112B}.

In this framework, a low value of $\alpha_{\rm mig}$ prevents outward migration, because the corotation torque saturates at low viscosities \citep{2001ApJ...558..453M}. On the other hand, the gap generated by the planet becomes deeper and thus inward migration becomes slower compared to higher values of $\alpha_{\rm mig}$ for large planets.

We thus investigate here just the influence of different migration speeds. This includes type-I migration, the transition from type-I to type-II migration as well as type-II migration itself. For this we vary $\alpha_{\rm mig}$ in section~\ref{sec:Kanagawamig}. Additionally, we keep $\alpha_{\rm disc}=0.0054$ through our simulations to have the same pebble scale heights in all simulations and thus the same pebble accretion rates.

As our simulations do not aim to model specific planetary systems with a specific migration history or to explain in detail observational data of exoplanets, we do not model the feedback from the planet on the protoplanetary disc, meaning that even if a planet is massive enough to generate a gap in the disc, the other planets still feel the unperturbed disc profile. However, we think this effect is not very important for the purpose of our study, because the width of the gap of a giant planet is roughly its Hill radius \citep{2006Icar..181..587C}, whereas planets in convergent migration are typically trapped in resonances outside of the Hill radius.

\section{Nominal migration rates}
\label{sec:nominal}

In this section we use our nominal planet migration recipe (see section~\ref{subsec:nominal} and \citealt{2015A&A...582A.112B}), where $\alpha_{\rm disc} = \alpha_{\rm mig} =0.0054$, and we ran 5 simulations for each setup with varying initial conditions (embryo mass, eccentricity, inclination and orbital elements).

\begin{figure*}
 \centering
 \includegraphics[scale=0.7]{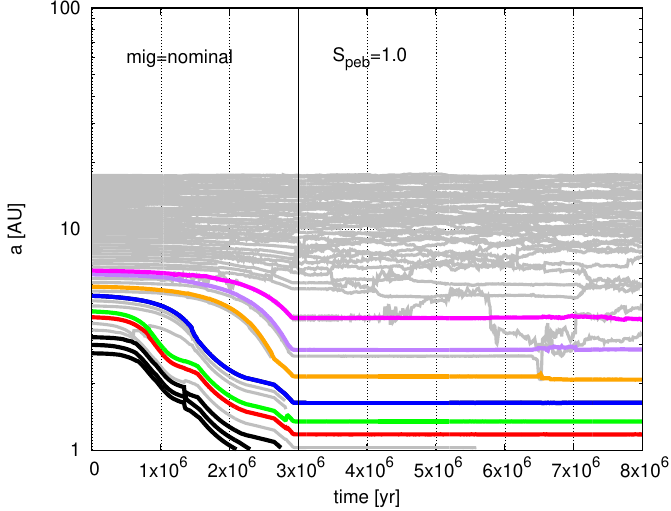}
 \includegraphics[scale=0.7]{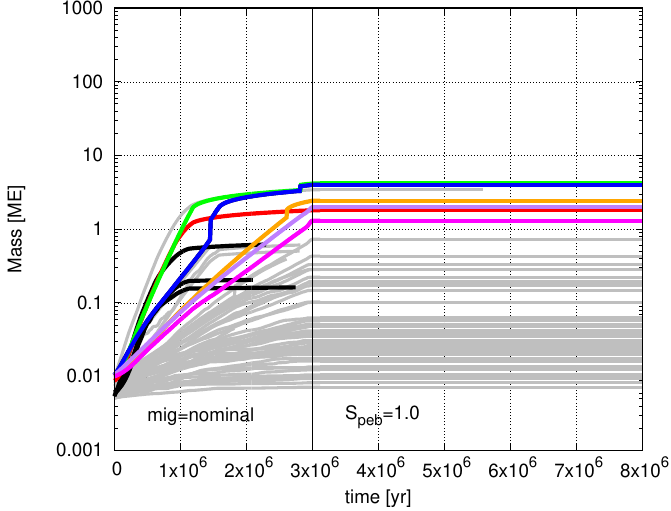}  
 \includegraphics[scale=0.7]{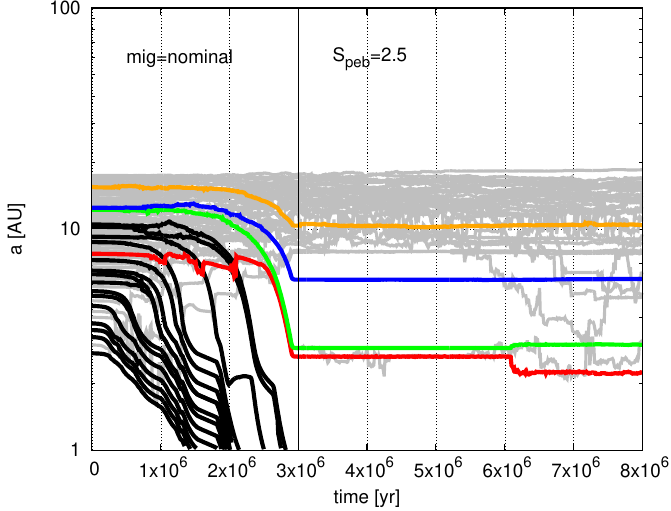}
 \includegraphics[scale=0.7]{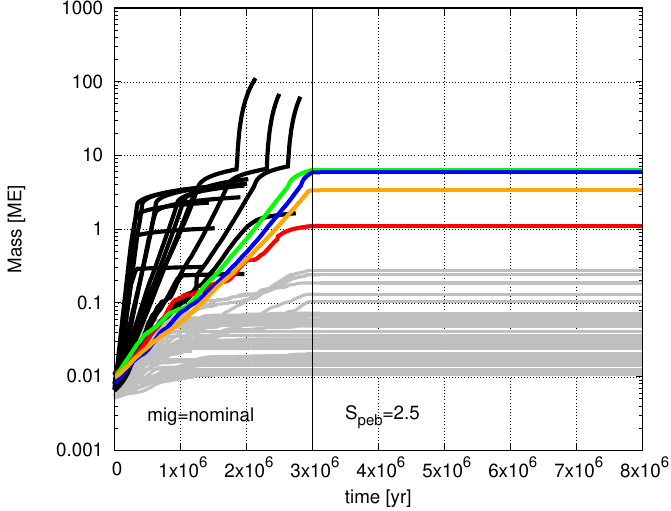}  
 \caption{Evolution of systems with nominal migration rate and two different pebble fluxes as function of time. Semi major axis (left) and planetary mass (right) of 60 planetary embryos with two different pebble scalings, $S_{\rm peb}=1.0$ (top) and $S_{\rm peb}=2.5$ (bottom). Bodies shown in black migrate by type-I migration interior to 1 AU, while grey bodies correspond either to small bodies that do not grow or to bodies that are ejected from the disc via gravitational interactions with other bodies. Coloured bodies correspond to the most massive surviving bodies in the disc. The gas disc lifetime is 3 Myr after injection of the planetary embryos, where the disc dissipation is marked by the vertical black line. The embryos migrate with nominal migration and the innermost embryo is placed at 2.75 AU.
   \label{fig:Nominalmigacc}
   }
\end{figure*}

We show in Fig.~\ref{fig:Nominalmigacc} the semi-major axis (left) and mass (right) evolution of 60 planetary seeds, where the innermost seed is placed at 2.75 AU. We use here the nominal pebble flux with $S_{\rm peb}=1.0$, meaning that 70 Earth masses of pebbles cross the disc during its lifetime and $S_{\rm peb}=2.5$ with a total of 175 Earth masses of pebbles. The value of $S_{\rm peb}$ is marked in each plot.

Even though the protoplanetary disc contains a region of outward migration that can trap bodies of up to a few Earth masses, planets migrate inwards. This is caused by two effects, (i) the region of outward migration is constrained to the inner disc (up to 2-3 AU, see \citealt{2015A&A...575A..28B}), so the inner bodies have to ``block`` the inward migrating outer bodies and might not be able to do so due to their low masses and (ii) due to mutual interactions between the bodies, the eccentricities increase, quenching the corotation torque responsible for outward migration \citep{2010A&A.523...A30, 2013A&A...553L...2C, 2014MNRAS.437...96F}. 

The net result is that planets migrate inwards and a few bodies cross towards the inner system with masses below 1 Earth mass. We note that in our simulations, planet migration becomes significant when a body reaches roughly 0.1 Earth masses. These bodies primarily originate from up to $\sim$5 AU. Planets forming exterior to 5 AU, on the other hand, grow to a few Earth masses and do not migrate interior to 1 AU. However, these bodies remain too small to start to accrete gas in an efficient way. 

Therefore, from the point of formation of gas giants, these sets of parameters fail. This is in agreement with the previous planet formation simulations by \citet{2015A&A...582A.112B} and \citet{2018A&A...609C...2B}, because the pebble flux is just too low to allow an efficient growth to core masses needed to accrete an gaseous envelope\footnote{We note that the nominal pebble flux used here corresponds to the red solid line of Fig.1 in \citet{2018A&A...609C...2B}. This pebble flux is too small to allow planets to grow to core masses needed to reach runaway gas accretion and results in a pebble surface density too low compared to observations.}.

Following the predictions of \citet{2018A&A...609C...2B}, an increase in the pebble flux will allow the growth of gas giants. Indeed using $S_{\rm peb}=2.5$ allows planetary embryos farther away to grow to reach pebble isolation mass earlier and before the end of the gas disc's lifetime, but they migrate into the inner disc. Planets might even reach runaway gas accretion, but they do not transition into type-II migration and thus migrate with the fast type-I rate into the inner regions of the disc (see appendix~\ref{ap:migration}).

The planets migrating inwards are slowed down in their migration in the region of outward migration caused by the entropy driven corotation torque, where the planets can form chains of resonant bodies that then migrate inwards as the disc slowly disperses. These chains of planets are very common at this evolutionary stage, because the planets grow to just a few Earth masses which is the correct planetary mass to be trapped in the region of outward migration\citep{2015A&A...575A..28B}. This effect was already observed in the N-body simulations of \citet{2017MNRAS.470.1750I} and can also be seen in our companion paper by \citet{Izidoro18}. For faster growth ($S_{\rm peb}=2.5$), the planets become too massive to be trapped in the region of outward migration and the whole chain migrates inwards and planets cross into the inner disc regions. Resonances for these outer systems are then not very common, but they can form chains of resonance anchored at the inner disc edge (see \citealt{2017MNRAS.470.1750I} and \citealt{Izidoro18}).

In Fig.~\ref{fig:Nominalmigacc}, the planetary systems shown are the most stable systems of this set of simulations. In fact, 80\% of our simulations using $S_{\rm peb}=1.0$ and 40\% of our simulations using $S_{\rm peb}=2.5$ become unstable after gas disc dispersal, even though the integration time is only 5 Myr after disc dispersal. This is related to the large number of planets with several Earth masses (and above) in a narrow region of $\sim$3 AU in the disc, where an outer belt of embryos with another couple of Earth masses can perturb the orbits of the inner planets. We note that the number of planets of at least a few Earth masses staying exterior to 1 AU is smaller for larger pebble fluxes, because the planets migrate inwards more efficiently (black lines in Fig.~\ref{fig:Nominalmigacc}). However, larger numbers of embryos at close distances to each other are easier to become unstable \citep{2006AJ....131.3093I, 2012Icar..221..624M}, explaining the differences in instabilities of the simulations with different pebble fluxes.

Additionally \citet{2018A&A...609C...2B} showed that gas giants staying outside of 1 AU must have originated from 20-40 AU. In the following we thus do not only vary the pebble flux, but also increase the orbital distance of the embryos in order to study the evolution in the outer disc to test if gas giant formation is possible in the outer disc within an N-body framework.

\begin{figure*}
 \centering
 \includegraphics[scale=0.66]{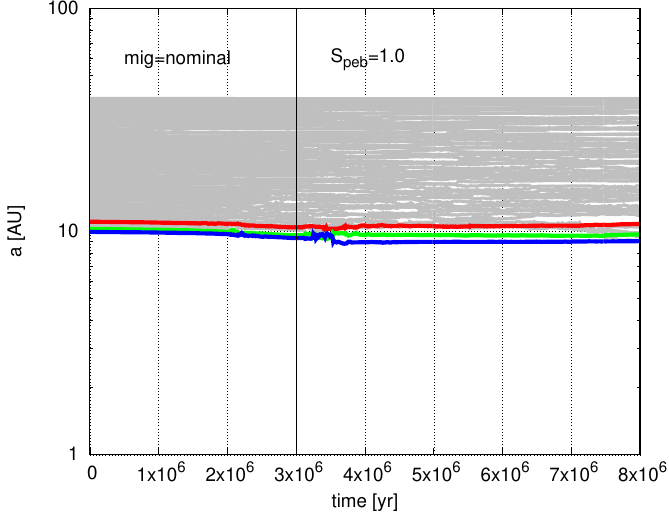}
 \includegraphics[scale=0.66]{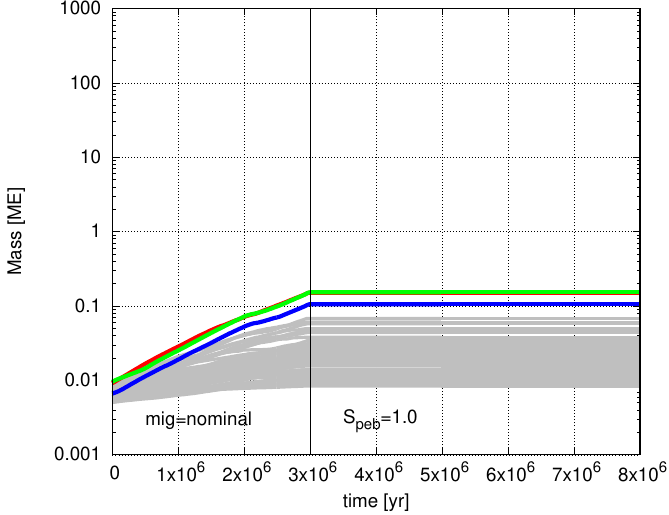}
 \includegraphics[scale=0.66]{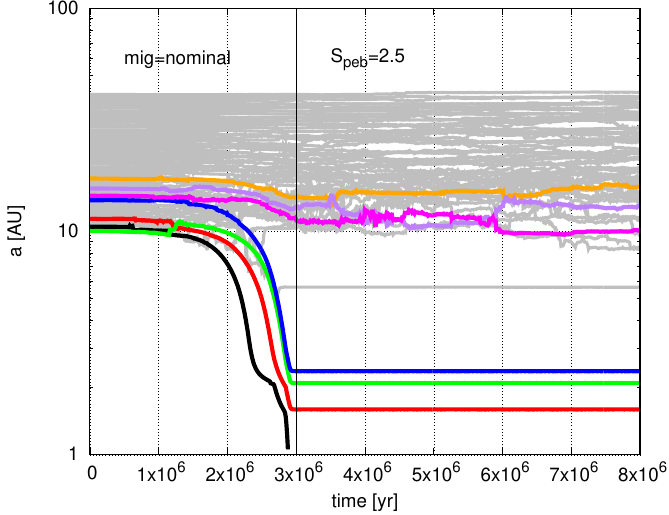}
 \includegraphics[scale=0.66]{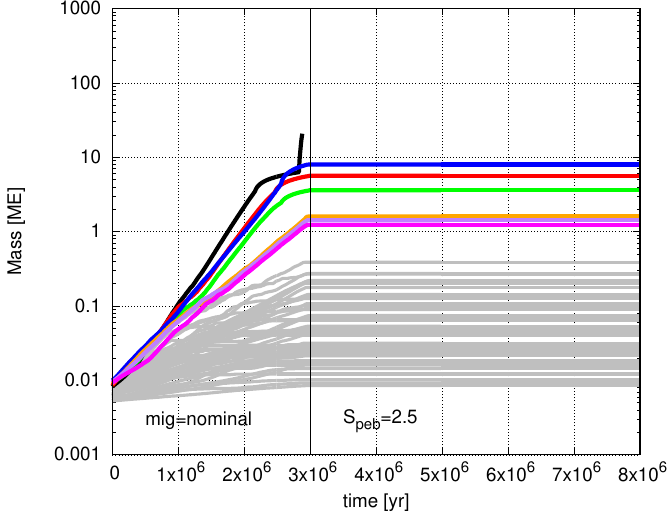}  
 \includegraphics[scale=0.66]{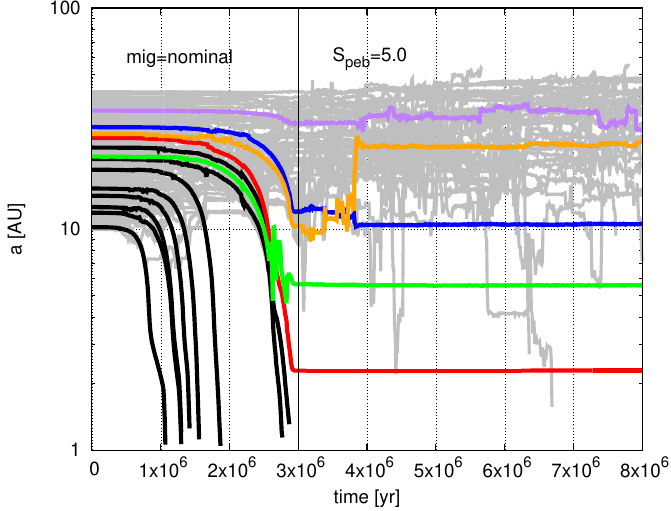}
 \includegraphics[scale=0.66]{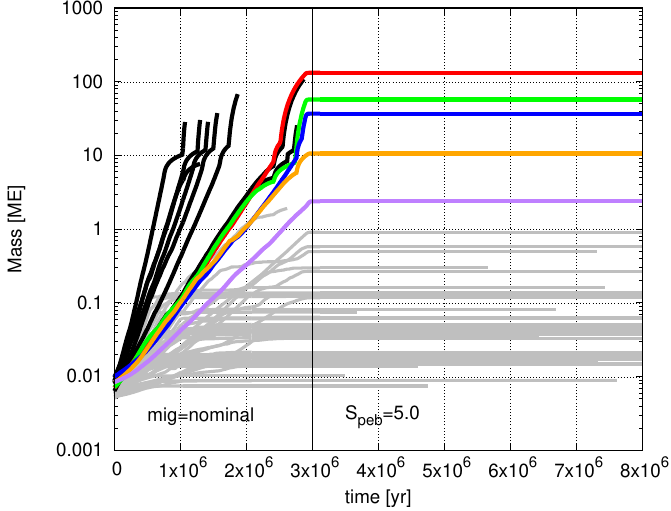} 
 \includegraphics[scale=0.66]{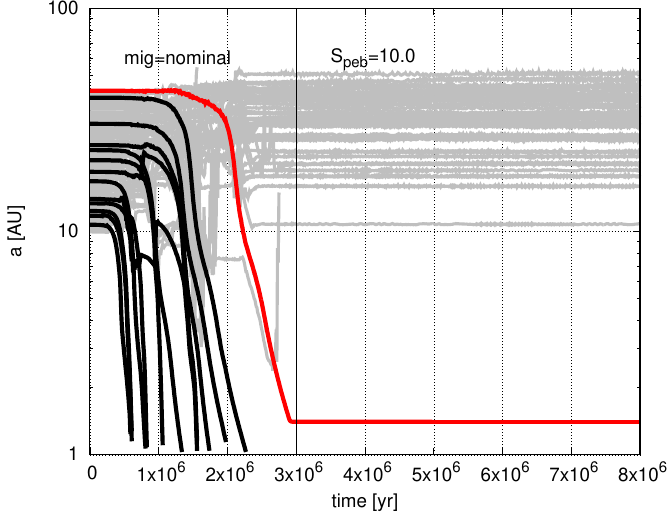}
 \includegraphics[scale=0.66]{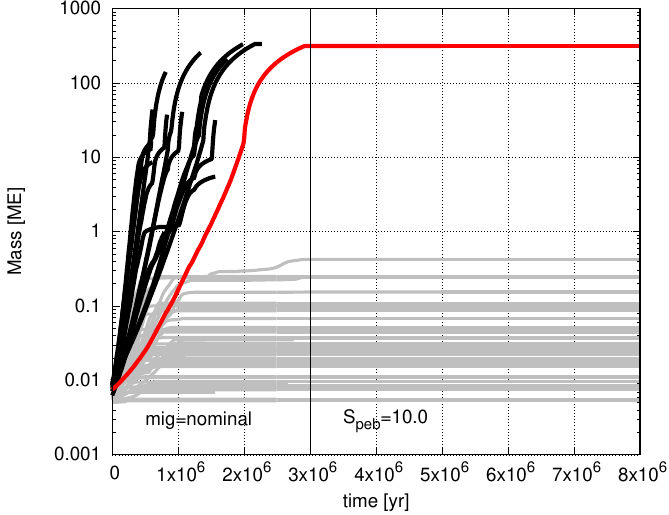}   
 \caption{Evolution of planetary systems with different pebble fluxes ($S_{\rm peb}=$ 1.0, 2.5, 5.0 and 10.0, top to bottom) and nominal migration rate in time. Semi major axis (left) and planetary mass (right) of 60 planetary embryos as function of time. The gas disc lifetime is 3 Myr after injection of the planetary embryos. Here the innermost embryo is placed at 10 AU, so further away than in Fig.~\ref{fig:Nominalmigacc}. As before, black lines correspond to bodies that migrate to the inner disc interior of 1 AU via type-I migration, bodies shown in grey are either not growing much or are ejected via gravitational interactions. The bodies shown in colours are the most massive surviving bodies in the simulation.
   \label{fig:largedistnominalmig}
   }
\end{figure*}

In Fig.~\ref{fig:largedistnominalmig} we show the semi-major axis (left) and planetary masses (right) of 60 planetary embryos in four different setups: $S_{\rm peb}=$ 1.0, 2.5, 5.0 and 10.0 (from top to bottom). The embryos are distributed from 10-40 AU. This means in the simulation with $S_{\rm peb}$=10.0, a total of 700 Earth masses in pebbles flow through the disc during the 3 Myr of evolution. This corresponds roughly to the suggested total amount of pebbles in \citet{2018A&A...609C...2B}.

As expected, the planetary embryos growing in the disc with the nominal pebble flux only show minimal growth and thus also only minimal planet migration. Their growth is very strongly hindered by the low pebble surface density in the outer disc.

Increasing the pebble flux by $S_{\rm peb}$=2.5 already allows the growth of some planetary embryos to masses that are big enough to reach gas accretion. Only the innermost embryos forming at around 10 AU grow quickly enough to reach these masses before gas disc dissipation. However these embryos migrate inwards close to 1 AU. For the embryos exterior to 15 AU, the pebble flux is still too low to grow significantly.

Increasing the pebble flux even further allows a more efficient growth, allowing seeds at larger orbital distances to grow to pebble isolation mass and eventually reach runaway gas accretion. Additionally, multiple seeds can grow in our simulations, resulting in about 10 planets that reach at least a few Earth masses, where actually most bodies reach pebble isolation mass. However, these planets are mostly lost to the inner disc, even when the bodies have already started to accrete gas in the runaway mode (see the black lines for $S_{\rm peb}=5.0$). This is due to the large viscosity ($\alpha=0.0054$), which only allows gap opening and transition to type-II migration when the planet is already very massive, so most planets do not reach gap opening mass at all in these sets of simulations.

As most of the planets migrate inwards very fast and as the planets are too massive to be trapped in the region of outward migration, planets do not get caught in resonances in the outer disc. The surviving planets for $S_{\rm peb}=5.0$ are also not in resonance and their final orbital positions are only determined by the gas disc dissipation. 

The giant planets in these simulations reach eccentricity of up to a few percent, which is caused by two effects: (i) as most giant planets are removed from the disc, not many massive bodies can interact to excite eccentricities and (ii) the damping due to interactions with the gas disc is quite efficient and eccentricities stay low \citep{2013A&A...555A.124B}. The smaller bodies (indicated in grey) have larger eccentricities, even above 10\%, similar to the grey lines in Fig.~\ref{fig:noedgenoscatter} and Fig.~\ref{fig:noedgescatter}.

In these simulations, instabilities after gas disc dispersal are not very common, only $\sim$5\% of our simulations show instabilities after gas disc dispersal. In case the growth is slow, the bodies that remain in the outer disc are so small that they do not excite large instabilities. The number of bodies that migrate inwards and survive exterior to 1 AU (red, green and blue lines for $S_{\rm peb}=2.5$ in Fig.~\ref{fig:largedistnominalmig}) is generally much smaller than in Fig.~\ref{fig:Nominalmigacc}, causing the systems to be more stable \citep{2012Icar..221..624M}. 

For the simulations with $S_{\rm peb}\geq5.0$, the planets grow and migrate fast into the inner disc, so that only a very small number of planets survive exterior to 1 AU. It might actually seem that giant planet growth is more efficient for $S_{\rm peb}=5.0$ than for $S_{\rm peb}=10.0$, however this is an artefact caused by the removal of bodies that cross interior to 1 AU (black lines). The bodies growing in discs with $S_{\rm peb}=10.0$ grow faster and thus migrate earlier so that more bodies can migrate into the inner disc until 3 Myr compared to planets formed in discs with $S_{\rm peb}=5.0$. In the case of $S_{\rm peb}=5.0$, planets just grow later and can thus survive in the outer disc. This can be seen by the time the first planet shown in black migrates interior to 1 AU. 

Figure~\ref{fig:largedistnominalmig} also shows that the innermost embryo grows first. This is related to the flaring of the disc structure with radius, resulting in a lower disc scale height at the position of the inner embryo, giving a lower $H_{\rm peb}$ and thus higher accretion rate. In any case, the growing planets leave the remaining embryos that grow slower on eccentric orbits preventing them from growing via pebble accretion. As the initial distances between the embryos are quite large, they remain stable \citep{2006AJ....131.3093I}.

Nevertheless, when $S_{\rm peb} \geq 5$, gas giants of Saturn mass and above form and can stay outside of 1 AU. The origin of these gas giants is beyond 30 AU as in \citet{2015A&A...582A.112B}. In fact, the Jupiter planet exterior to 1 AU formed in the simulation with $S_{\rm peb}=$10 originates from around 40 AU. As it migrates through the disc, it scatters away the remaining embryos on eccentric orbits, quenching their growth.

These simulations confirm the single body simulations of \citet{2015A&A...582A.112B} and \citet{2018A&A...609C...2B} in the sense that the formation of gas giants outside of 1 AU is possible, but only if the planetary embryos to form the gas giants originate from the outer disc at 20-40 AU. The total pebble mass has to be at least 300 Earth masses in our simulations to allow the formation of gas giants exterior to 1 AU. Planetary embryos formed interior to 30 AU still grow, but migrate to the inner disc with a variety of masses (from 0.1 Earth masses to even above Saturn mass) and would eventually disrupt an inner planetary system \citep{2005A&A...441..791F, 2006Sci...313.1413R, 2007ApJ...660..823M}. Some of these bodies would then grow to become hot Jupiters as predicted by \citet{2015A&A...582A.112B} and \citet{2018A&A...609C...2B}, where the inner disc edge stops their inwards migration before they migrate all the way into the central star. In \citet{Izidoro18} these fast migration rates are used to reproduce systems of super Earths. In fact, instabilities of these super Earths can produce planetary cores of several 10 Earth masses. We note however that we do not observe early instabilities and collisions between the bodies and growth up to several Earth masses is entirely dominated by pebble accretion.

These results make it very difficult to explain the formation of the Solar system, if planetary embryos form all over the disc. Our model of planet growth suggests two possible scenarios to avoid the invasion of inner planetary systems from bodies growing in the outer system, (i) the planetary seeds are not distributed all over the disc, but features certain pile-ups, for example at ice lines \citep{2017A&A...602A..21S, 2017A&A...608A..92D} to avoid too many seeds to grow to several Earth masses. However, if the seeds form too close, they would still migrate into the inner system unless events like outward migration in resonance of giant planets are invoked \citep{2001MNRAS.320L..55M, 2014ApJ...795L..11P, 2011Natur.475..206W} or (ii) that planet migration is slower than anticipated in our nominal model. We will focus in the following to study the effects of reduced migration rates on the formation of planetary systems.

\section{Reduction of migration}
\label{sec:Kanagawamig}

We now investigate the influence of different migration speeds set by $\alpha_{\rm mig}$ on the formation of planetary systems. Just reducing the type-II migration rate alone will not keep the growing planets exterior to 1 AU, because gap opening in high viscosity discs only happens at very large planet masses (see appendix~\ref{ap:migration}).

We do not alter the growth rates and pebble isolation mass, meaning that $H_{\rm peb}$ and the disc structure are determined by $\alpha_{\rm disc}=0.0054$. In appendix~\ref{ap:Hpeb} we present simulations with a reduced $H_{\rm peb}$, which essentially reproduce planetary systems formed in discs with higher $H_{\rm peb}$ and higher $\dot{M}_{\rm peb}$.

We remind the reader that the positive torques, responsible for outward migration driven by the entropy related corotation torque, saturate for low viscosities. This means that if $\alpha_{\rm mig}$ is reduced outward migration might not be possible any more. Even though outward migration is possible in the previous simulations, planets mostly migrate inwards due to their eccentricities being larger than the horseshoe width of the planet \citep{2010A&A.523...A30}. On the other hand, a reduced $\alpha_{\rm mig}$ parameter will allow a faster gap opening and thus slower migration, if the planet grows big enough. We test two variations of $\alpha_{\rm mig}$ with $0.00054$ and $0.0001$, so a factor of 10 and $\sim$50 smaller than $\alpha_{\rm disc}$. We ran 5 simulations for each setup with varying initial conditions (embryo mass, eccentricity, inclination and orbital elements).

\begin{figure*}
 \centering
 \includegraphics[scale=0.66]{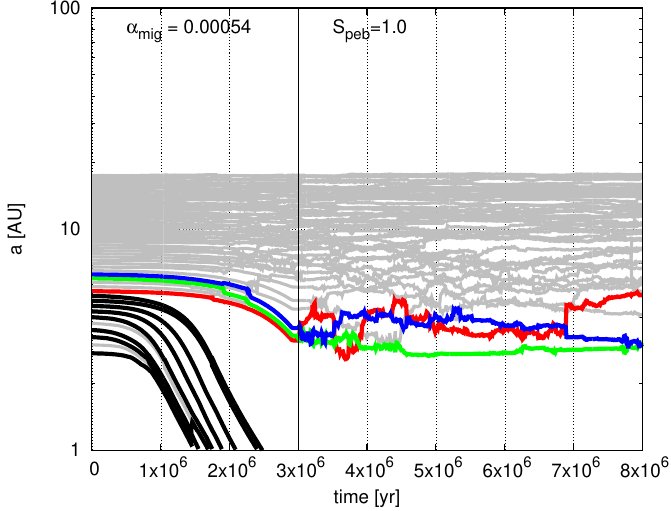}
 \includegraphics[scale=0.66]{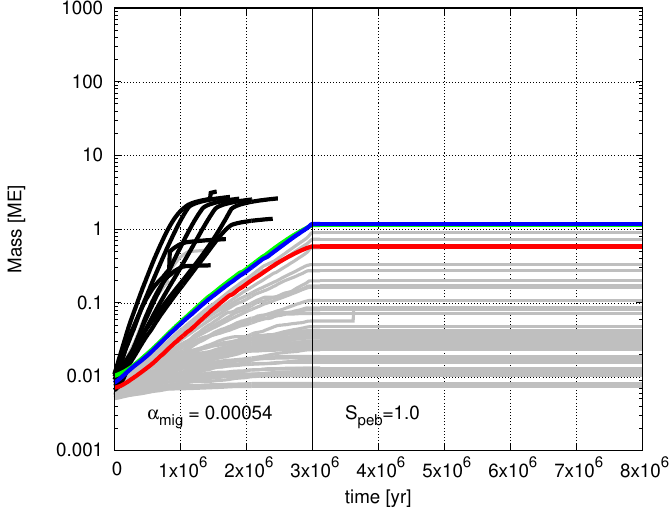}
 \includegraphics[scale=0.66]{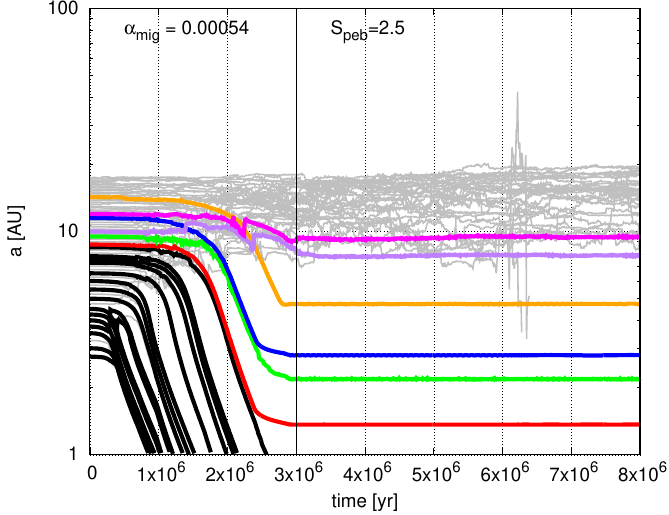}
 \includegraphics[scale=0.66]{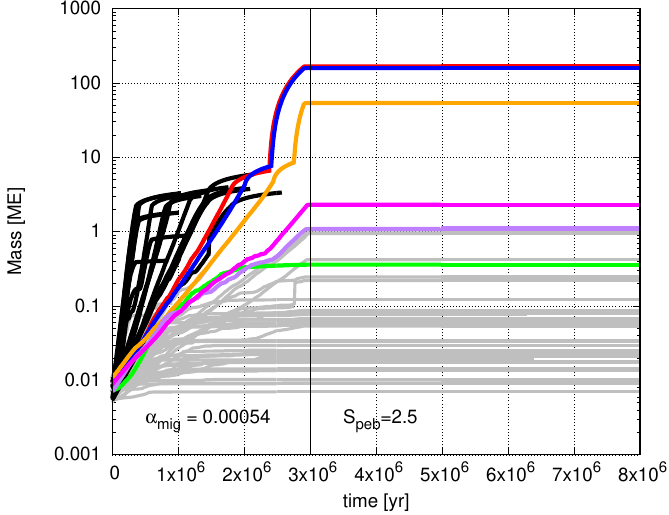}  
 \includegraphics[scale=0.66]{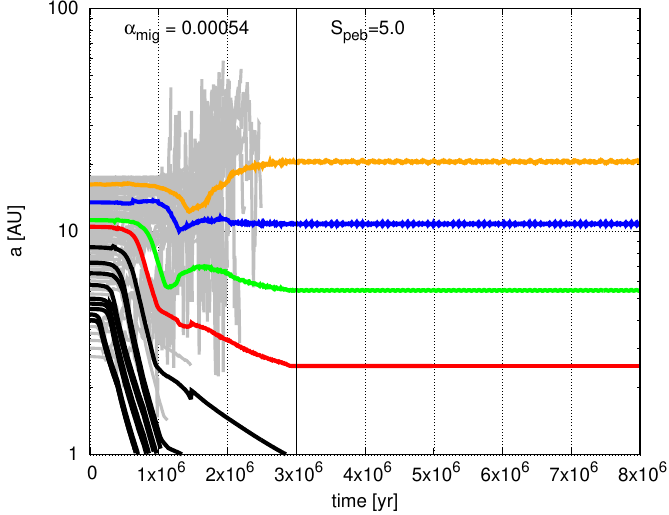}
 \includegraphics[scale=0.66]{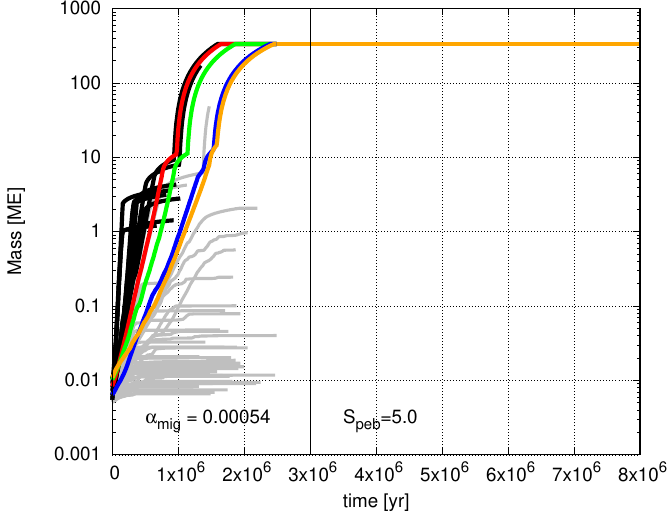} 
 \includegraphics[scale=0.66]{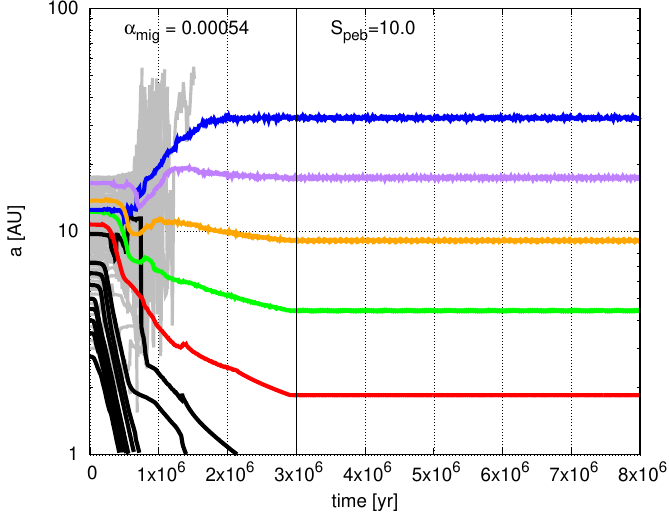}
 \includegraphics[scale=0.66]{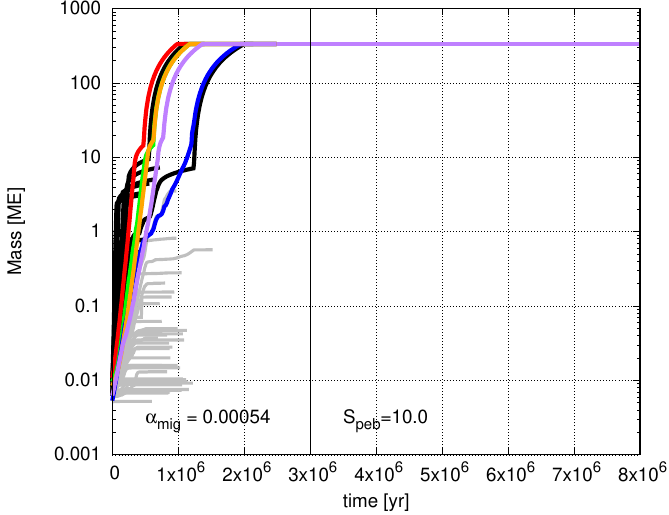}   
 \caption{Evolution of planetary systems with different pebble fluxes ($S_{\rm peb}=$ 1.0, 2.5, 5.0 and 10.0, top to bottom) and $\alpha_{\rm mig}=0.00054$ in time. Semi major axis (left) and planetary mass (right) of 60 planetary embryos as function of time. The gas disc lifetime is 3 Myr after injection of the planetary embryos. The black, grey and coloured lines show the same behaviour as bodies in Fig.~\ref{fig:largedistnominalmig}.
   \label{fig:Kanagawamig00054}
   }
\end{figure*}

In Fig.~\ref{fig:Kanagawamig00054} we show the evolution of 60 planetary seeds in discs with four different pebble fluxes of $S_{\rm peb}=$1.0, 2.5, 5.0 and 10.0, where $\alpha_{\rm mig}=0.00054$. Otherwise the simulation parameters (disc structure, evolution, etc.) are the same as in the previous simulations (Fig.~\ref{fig:Nominalmigacc}). We remove again bodies that cross interior to 1 AU to save computation time.

Clearly a reduction of $\alpha_{\rm mig}$ compared to Fig.~\ref{fig:Nominalmigacc} reduces the distance the planets migrate through the disc during the gas disc's lifetime (see also appendix~\ref{ap:migration}). Due to the reduced migration speed, planetary embryos forming at around 10 AU can stay outside of 1 AU, if the pebble flux allows the formation of gas giants. In the case of the nominal pebble flux, planets can grow to a few Earth masses, if they form interior to 5 AU, but then they migrate interior of 1 AU. For low mass planets, the reduced $\alpha_{\rm mig}$ parameter prevents the torques to stay unsaturated and thus allows only inward migration. This actually implies that the formed low mass planets migrate to the inner disc faster than in the scenario with nominal migration (see Fig.~\ref{fig:Nominalmigacc}). However, the reduced $\alpha_{\rm mig}$ allows a transition to slower type-II migration if the planets become big enough.

Comparing the simulations with nominal pebble flux to the simulations with $S_{\rm peb}=2.5$, planetary embryos from within 5 AU grow to similar masses. This is because planet growth is limited by the pebble isolation mass, which is so small that planets do not go into runaway gas accretion (appendix~\ref{ap:Miso}).

In the cases of even larger pebble fluxes, more giant planets form from the planetary embryos. However, only the giant planets formed from seeds growing beyond 10 AU stay exterior to 1 AU. Planets formed interior to 10 AU still cross the 1 AU line, because the migration rates for those planets are still too fast. Following the planet's evolution indicated by the black line crossing into 1 AU at $\approx$2.8 Myr in the simulations with $S_{\rm peb} = 5.0$ indicates that the regime after gap opening still accounts for a significant reduction of semi-major axis.

Due to the low viscosity, planets that reach runaway gas accretion migrate very slow and can act as barrier for planets migrating inwards from the outer disc, similar as invoked in the scenario by \citet{2015A&A...582A..99I}. Bodies migrating faster than the giants could be trapped in mean motion resonances, however, most of the inward migrating planets also grow to gas giants and migrate slow, so that they do not come close enough to reach mean motion resonances (planets stay exterior to the 2:1 resonance). In fact, the giants shown by the red and blue lines in the $S_{\rm peb}=2.5$ case in Fig.~\ref{fig:Kanagawamig00054} are close to the 3:1 MMR, but not in it. This is simply caused by the transition of the planet shown in blue into type-II migration by its growth before it comes close to the resonance. Also for the cases with higher pebble fluxes ($S_{\rm peb}\geq5.0$) we do not observe planets being trapped in resonances. This is enhanced by the similar migration speeds of the planet due to their similar masses\footnote{If planets migrate with the same speed, they can not be trapped in resonance, which is why the simulations by \citet{2017A&A...598A..70S} invoked that the innermost planet does not migrate at all to allow trapping in resonance in the first place.}.

The outward migration of the giant planets in the outer disc for $S_{\rm peb}=5.0$ and $S_{\rm peb}=10.0$ is caused by the interactions between the massive planets. In the case of $S_{\rm peb}=5.0$, a small instability occurs at 1.5 Myr, visible by the small jump in semi-major axis for the planets shown in black and red. This instability is caused by the ejected of a 50 Earth mass planet (shown in grey), which also coincides with the outward migration of the orange planet. A similar event is observed for the $S_{\rm peb}=10.0$ simulation, when the planet shown in black loses a lot of its semi-major axis at 800kyr.

Clearly, a reduction of $\alpha_{\rm mig}$ pushes the potential formation locations of gas giants to the inner disc. However, if planetary seeds form all over the disc, our simulations indicate that this scenario is still inconsistent with the Solar System due to the inward migration of gas giants interior to 1 AU. In the following, we thus study the effects of an even smaller $\alpha_{\rm mig}$ parameter.

\begin{figure*}
 \centering
 \includegraphics[scale=0.66]{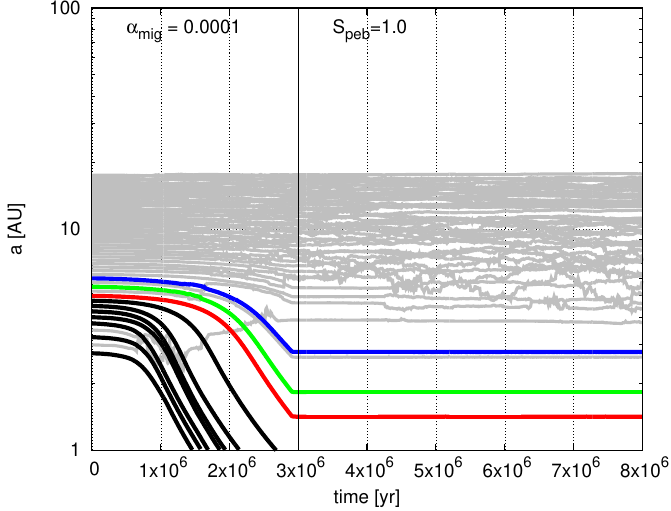}
 \includegraphics[scale=0.66]{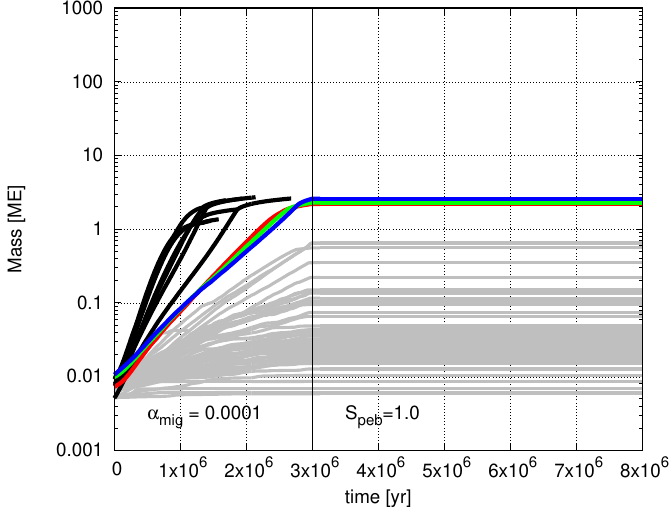}
 \includegraphics[scale=0.66]{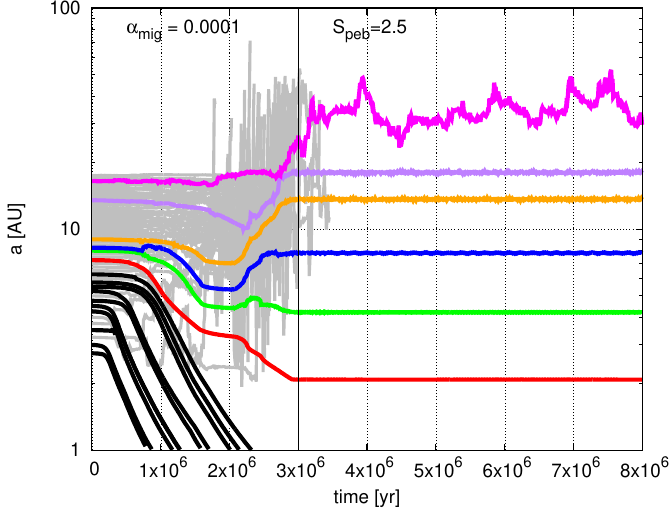}
 \includegraphics[scale=0.66]{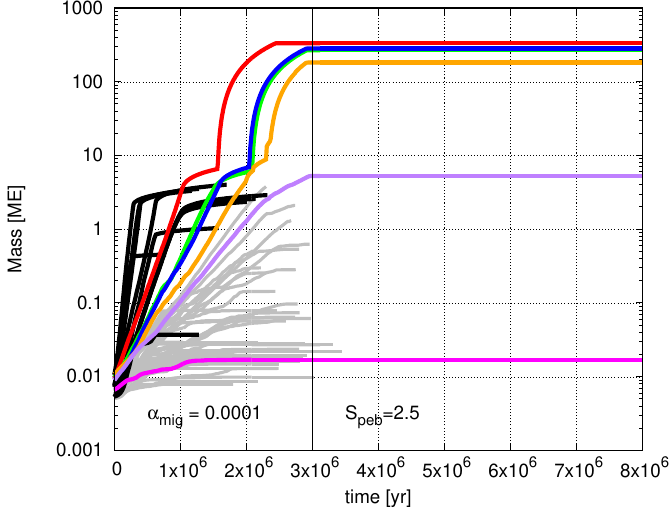}  
 \includegraphics[scale=0.66]{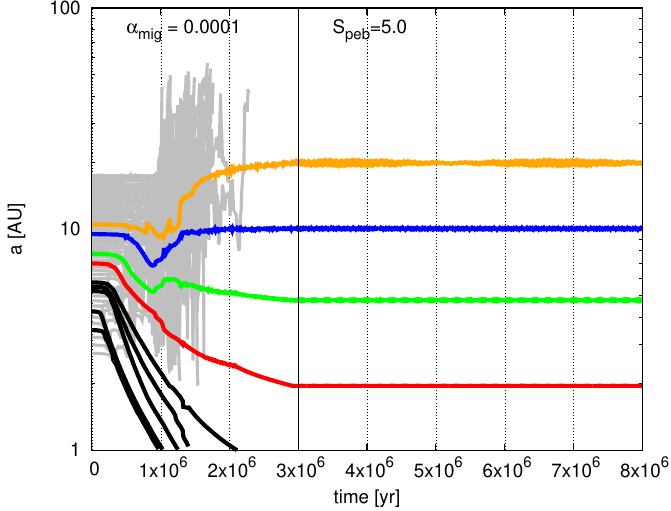}
 \includegraphics[scale=0.66]{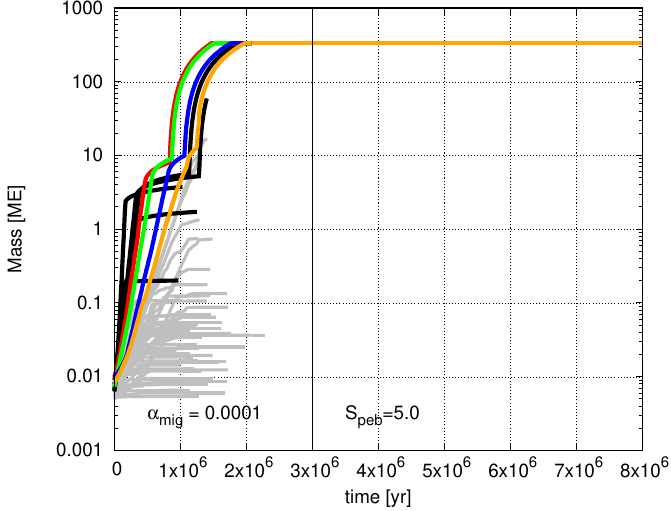} 
 \includegraphics[scale=0.66]{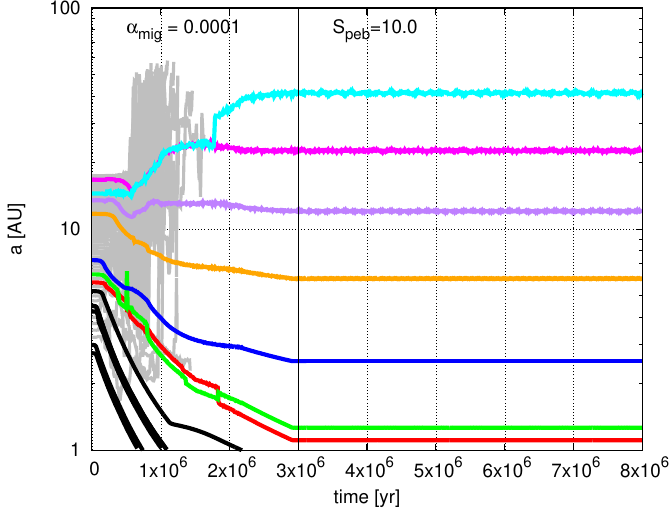}
 \includegraphics[scale=0.66]{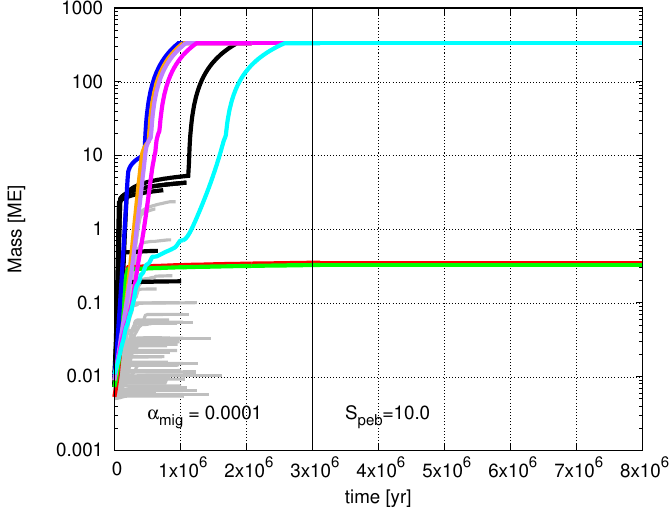}   
 \caption{Evolution of planetary systems with different pebble fluxes ($S_{\rm peb}=$ 1.0, 2.5, 5.0 and 10.0, top to bottom) and $\alpha_{\rm mig}=0.0001$ in time. Semi major axis (left) and planetary mass (right) of 60 planetary embryos as function of time. The gas disc lifetime is 3 Myr after injection of the planetary embryos. The black, grey and coloured lines show the same behaviour as bodies in Fig.~\ref{fig:largedistnominalmig}.
   \label{fig:Kanagawamig0001}
   }
\end{figure*}

In Fig.~\ref{fig:Kanagawamig0001} we show the evolution of 60 planetary seeds in discs with four different pebble fluxes, where $\alpha_{\rm mig}=0.0001$. Otherwise the simulation parameters (disc structure, etc.) are the same as in the previous simulations. In the case of the nominal pebble flux even the lower migration rates compared to the simulations shown in Fig.~\ref{fig:Nominalmigacc} can not keep the planets exterior to 1 AU long enough to grow large enough to be able to trigger gas accretion.

Additionally due to the low $\alpha_{\rm mig}$, outward migration driven by the entropy related corotation torque is quenched, resulting in a faster inward migration of the very innermost planetary embryos as for $\alpha_{\rm mig}=0.0054$. In fact, only planets interior of 5 AU start to grow efficiently in this case similar to the simulations using $\alpha_{\rm mig}=0.0054$ and the nominal migration rates (Fig.~\ref{fig:Nominalmigacc}).

However, using $S_{\rm peb}$=2.5 allows the growth of gas giants. Due to the higher pebble flux, planets exterior to 5 AU can start to grow efficiently as well. In fact, the embryos outside of $\approx$5 AU are the ones that grow to become gas giants in all our simulations with $\alpha_{\rm mig}=0.0001$. This is caused by two effects, (i) the disc's aspect ratio is larger in the outer disc, which allows a larger pebble isolation mass and thus larger core masses making the transition into runaway gas accretion easier (appendix~\ref{ap:Miso}) and (ii) the planetary embryos growing interior of 5 AU have to cover less distance to reach 1 AU making them more prone to migrate interior to 1 AU. This effect is visible in all simulations with $S_{\rm peb}\geq 2.5$.

In all simulations with $S_{\rm peb} \geq 2.5$ at least 3-4 gas giants form, where the number of gas giants increases with pebble flux. Relating the pebble flux to the metallicity of the system our simulations are in agreement with the metallicity correlation for giant planets \citep{2005ApJ...622.1102F, J2010}, where the occurrence rate of giant planets increases with host star metallicity. Additionally, the formation of multiple giant planets per systems increases the probability that scattering events might occur, leading to an eccentricity distribution like that of observed giant planet, which increases with host star metallicity as well \citep{2013ApJ...767L..24D, 2018arXiv180206794B}.

However, in all simulations planets of at least a few Earth masses migrate into the inner discs, making these simulations not conform with the solar system structure, even if gas giants can stay exterior of 1 AU. We discuss more about planets migrating into the inner system and long term simulations in section~\ref{sec:systems} and section~\ref{sec:disc}. Similar to the simulations with $\alpha_{\rm mig}=0.00054$, thee giant planet in our systems are close to some high order mean motion resonances (e.g. 1:3, 2:7), but not in it due to the same effects.


Interestingly, only in the simulations with $S_{\rm peb} = 2.5$ we see the formation of an ice giant in the outer disc. In simulations with larger pebble flux, the growth timescale is so short that the planets reach pebble isolation mass in the outer disc and grow to become gas giants. This is consistent with the simulations of \citet{2014A&A...572A..35L} and \citet{2015A&A...582A.112B}, where ice giants are only formed in a small area of parameter space. They either grow too fast and become gas giants or grow too slow and do not grow at all. This trend holds for both $\alpha_{\rm mig}$ assumptions. In principle the ice giants could form by giant impacts from planets of a few Earth masses exterior to the gas giants \citep{2015A&A...582A..99I}, however, we also very rarely observe bodies of this mass in the outer disc in our simulations. This could be related to the distribution and masses of our planetary embryos (see section~\ref{sec:disc}).

\begin{figure}
 \centering
 \includegraphics[scale=0.7]{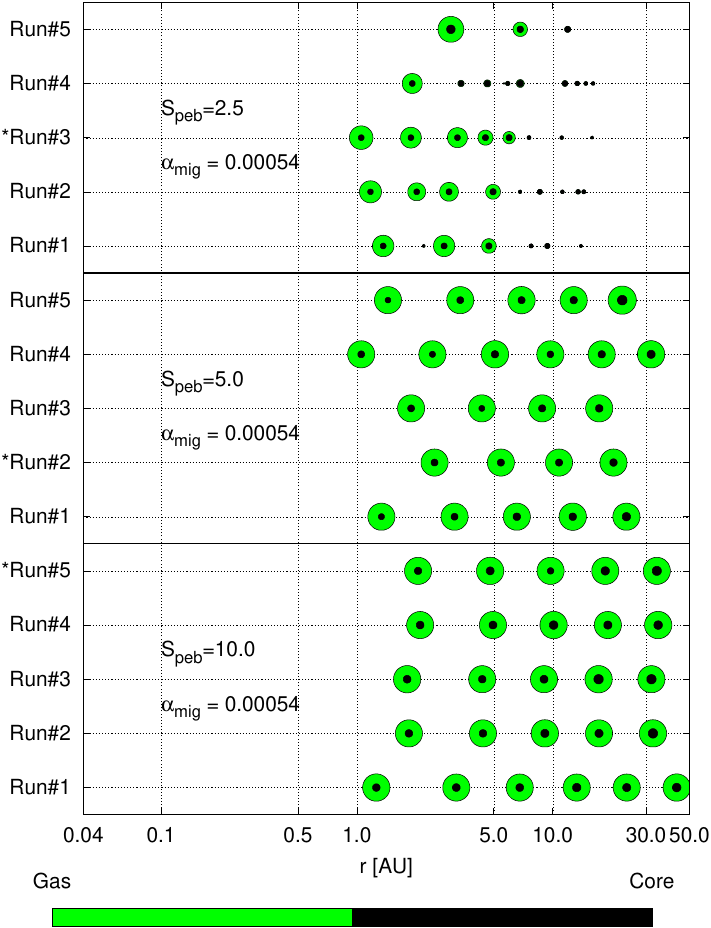}
 \includegraphics[scale=0.7]{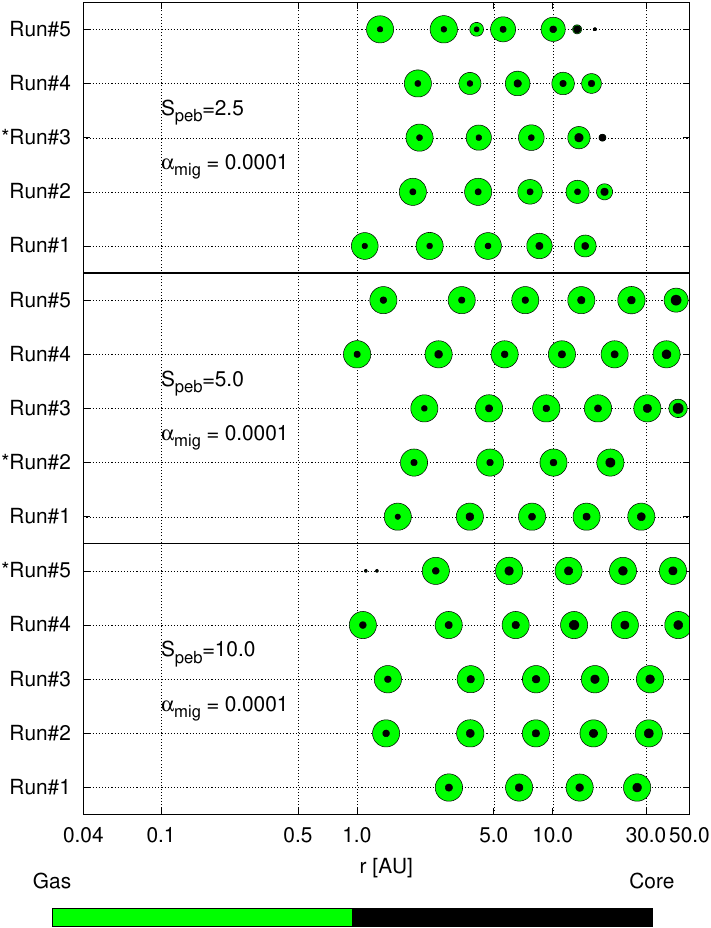}
 \caption{Final configurations after 8 Myr of integration of all our simulations with $\alpha_{\rm mig}=0.00054$ (top) and $\alpha_{\rm mig}=0.0001$ (bottom) for different pebble fluxes as marked in the figure. Planets migrating interior to 1 AU are removed from the simulations here. We only show planets with masses above 0.3${\rm M}_{\rm E}$. The size of the circle is proportional to the total planetary mass (green) by the 3rd root and to the mass of the planetary core (black) also by the 3rd root. The runs marked with * are the example runs shown in Fig.~\ref{fig:Kanagawamig00054} and Fig.~\ref{fig:Kanagawamig0001}.
   \label{fig:Sim00054}
   }
\end{figure}

We present in Fig.~\ref{fig:Sim00054} the final configurations of the different planetary systems formed with $\alpha_{\rm mig}=0.00054$ (top) and $\alpha_{\rm mig}=0.0001$ (bottom) for pebble fluxes with $S_{\rm peb} \ge 2.5$. Each run differs only by varying the initial masses, eccentricities, inclinations and orbital parameters of the planetary embryos within our specified setup. We only show planets with masses larger than 0.3${\rm M}_{\rm E}$.

Using $S_{\rm peb}=2.5$, a lower $\alpha_{\rm mig}$ seems to result in the formation of more gas giants in the outer disc and less small bodies.  For slow migration, the growing planets stay embedded in their starting environment of planetary embryos and they scatter their smaller counterparts, preventing them from accreting. At the end of the gas discs lifetime these small bodies are ejected leaving only the big bodies behind. For $\alpha_{\rm mig}=0.00054$, the planets migrate faster and thus leave their birth location. As the growing planet has migrated inwards, the small embryos are too far away to be influenced by the formed gas giants and thus stay in the disc until the end of our simulations where they only grow moderately.

This implies that in our model, two requirements need to be fulfilled for planetary embryos to grow to gas giants, (i) the pebble flux needs to be large enough for planetary cores to grow to pebble isolation mass, where a minimum flux of 170 Earth masses over the disc lifetime is need and (ii) the planetary embryos need to form far enough away from the central star to achieve a pebble isolation mass large enough to allow a contraction of the gaseous envelop (see appendix~\ref{ap:Miso}). For the formation of gas giants exterior to 1 AU, the migration rate also needs to be slow enough to keep the planets in the outer disc.

\section{Mass and composition of the final systems}
\label{sec:systems}

\begin{figure}
 \centering
 \includegraphics[scale=0.7]{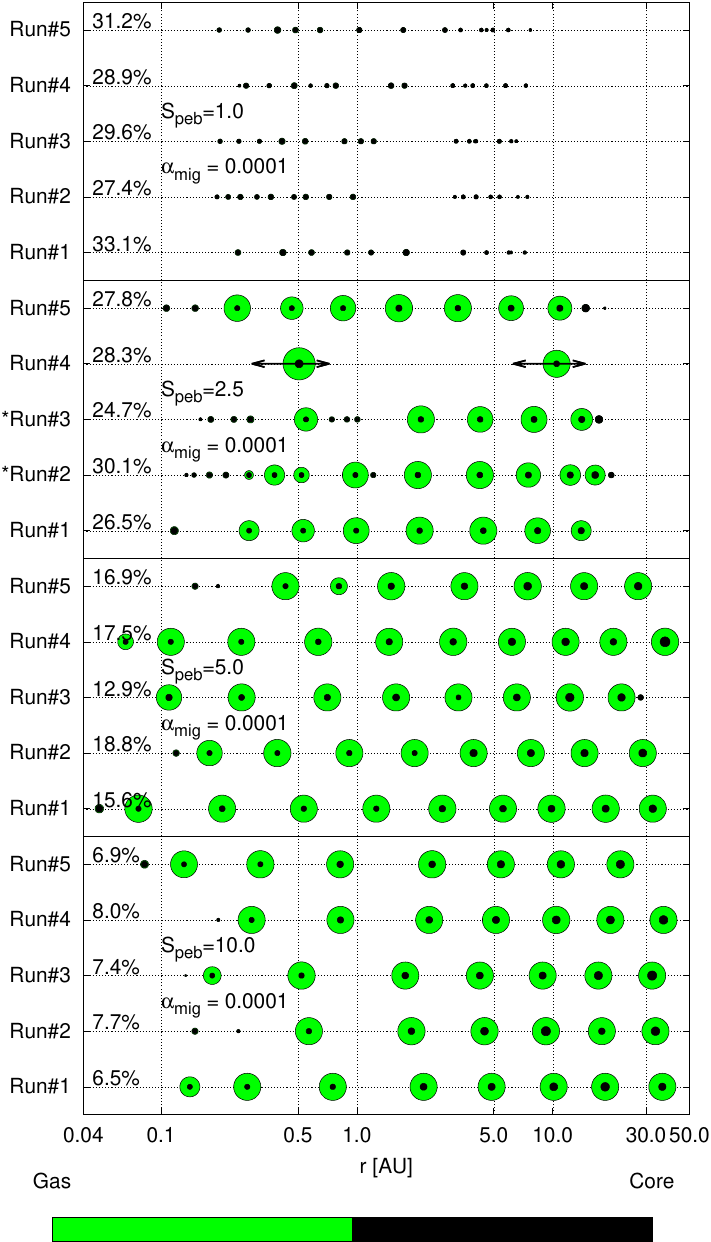}
 \caption{Final configurations after 8 Myr of integration of all our simulations with $\alpha_{\rm mig}=0.0001$ and different pebble fluxes. The size of the circle is proportional to the total planetary mass (green) by the 3rd root and to the mass of the planetary core (black) also by the 3rd root. The runs marked with * are the example long-term runs shown in Fig.~\ref{fig:noedgenoscatter} and Fig.~\ref{fig:noedgescatter}, while the black arrows indicate the aphelion and perihelion positions of the planet calculated through $r_{\rm P} \pm e\times r_{\rm P}$. The percentage numbers in front show the pebble-to-planet conversion ratio $f_{\rm p2p}$, which is discussed in section~\ref{sec:disc}.
   \label{fig:noedge0001}
   }
\end{figure}

We now show example outcomes of the planetary systems formed from our simulations. In order to capture the whole structure of the planetary systems, planets are now allowed to migrate into the inner disc and are not removed from the simulation as in the previous sections, increasing the computational time by a factor of 50-100 due to the tight inner orbits that planets can reach. We focus here on simulations with $\alpha_{\rm mig} = 0.0001$ and different pebble fluxes and show the systems after 8 Myr of evolution in Fig.~\ref{fig:noedge0001}.

The systems formed with the nominal pebble flux $S_{\rm peb}=1.0$ show systems of multiple super-Earth planets (of up to a few Earth masses). These super-Earths actually do not reach the inner edge of the disc at 0.1 AU, because they migrate very slowly caused by them being small during a significant fraction of the disc lifetime. This is in contrast to our companion papers \citep{Izidoro18, Lambrechts18} where planets reach the inner edge of the disc. This difference is caused by the different starting positions of the planetary seeds, which are close to the host star in our companion papers. Additionally, the migration speed in itself differs between our papers. \citet{Lambrechts18} only considers inward migration, while \citep{Izidoro18} uses the migration rate that corresponds to our nominal case (section~\ref{sec:nominal}), whereas the simulations presented in Fig.~\ref{fig:noedge0001} use $\alpha_{\rm mig}=0.0001$.

In our companion papers, the super-Earths are driven to to the inner edge of the disc, where they can be trapped in resonance chains that eventually become unstable after gas disc dissipation. These effects and how the resulting systems of super-Earths match to observations are discussed in our companion paper by \citet{Izidoro18}.

As in the simulations where planetary embryos were removed when crossing interior to 1 AU, the formed planetary systems harbour many gas giants, when $S_{\rm peb} \ge 2.5$. However, in the inner systems, super-Earth planets can have formed also for higher pebble fluxes. The fraction of super-Earth planets in the inner disc is lower compared to $S_{\rm peb}=1.0$ and decreases with increasing $S_{\rm peb}$. In fact for increasing $S_{\rm peb}$, the systems with inner hot super-Earths decreases.

The reason why slightly fewer super-Earths form in the systems with higher pebble flux is related to the faster growth of the planets. At higher pebble flux, planets reach pebble isolation mass earlier and have thus more time to contract their gaseous envelope and go into runaway gas accretion. This especially applies to the inner disc, where the pebble isolation mass is small (appendix~\ref{ap:Miso}) and longer times for envelope contraction are needed.


However, the innermost gas giants in the runs with larger pebble flux do not reach Jupiter mass until the end of the disc's lifetime. As runaway gas accretion is only limited to the accretion rate of the protoplanetary disc in our simulations, gas giants that did not reach Jupiter mass must have started to accrete gas in the runaway gas accretion regime at very late stages. This implies that the faster core growth allows the transition into runaway gas accretion for these bodies. Our simulations thus imply that giant planets originating from the inner disc should be less massive than their counterparts formed in the outer disc and harbour smaller planetary cores due to the smaller pebble isolation mass. This effect is clearly visible in Fig.~\ref{fig:noedge0001}, where the cores of the inner gas giants are smaller than for the outer gas giants.

The core masses of the innermost bodies are in the range of $\approx$4 Earth masses, where it is still debated if this mass is large enough to trigger runaway gas accretion (see section~\ref{sec:disc}). Longer contraction time-scales in contrast to our simulations would thus prevent the growth of these super-Earth planets to gas giants. Observations of exoplanets show a super-Earth occurrence rate independently of host star metallicity \citep{2012Natur.486..375B}, implying that our gas accretion rates are overestimated due to the failure of producing super Earth systems for high pebble fluxes. On the other hand, super Earths might still be produced if planetary seeds start even closer to their host star.

Our simulations also clearly show that the formation of super-Earths and gas giants is not mutually exclusive. In fact, our simulations with $S_{\rm peb} = 2.5$ show structures with close-in super-Earths, exterior gas giants and even further populations of super-Earths and gas giants all in the same system (see also Fig.~\ref{fig:noedgenoscatter}). This is in line with new observational data \citep{2018arXiv180502660Z, 2018arXiv180608799B} and we discuss this more in section~\ref{sec:disc}.

This structure seems at first in contradiction with the simulations by \citet{2015ApJ...800L..22I}, who showed that giant planets can block inward migrating super-Earths. They assumed that a gas giant grew inwards of any other forming planets. Indeed we observe the same behaviour that a growing gas giant (indicated by the magenta line in Fig.~\ref{fig:noedgenoscatter}) blocks the inward migration of exterior super-Earths (shown by the three black lines exterior to the magenta line). The scenario invoked in \citet{2015ApJ...800L..22I} therefore remains valid, namely that gas giants can block exterior planets, but super-Earth planets can form also interior of gas giants as shown in our simulations. 

However, the formed systems are highly packed and could thus become unstable. We therefore investigate the long-term evolution of these systems in the following.

Instabilities are thought to be the norm in exoplanet systems. Gravitational perturbations cause the orbits of giant planets to cross, which leads to a phase of close encounters and planet-planet scattering. These instabilities generally culminate when one or more planets are ejected from the system, and the surviving planets have eccentric orbits \citep{1996Sci...274..954R, 1996Natur.384..619W, 1997ApJ...477..781L, 2008ApJ...686..621F}. Matching the observed eccentricity distribution of giant exoplanets requires that at least 75\% -- and probably 75-90\% -- of giant planet systems observed today are the survivors of instabilities \citep{2005Icar..178..517M, 2008ApJ...686..603J, 2010ApJ...711..772R}. Giant planet instabilities also have a profound effect on the other parts of their planetary systems, both by disrupting terrestrial planets (or their progenitors) and outer planetesimal belts \citep{2006ApJ...645.1509V, 2011A&A...530A..62R, 2012A&A...541A..11R}, as we confirm in our simulations below.

\subsection{Individual systems}

\begin{figure*}
 \centering
 \includegraphics[scale=0.7]{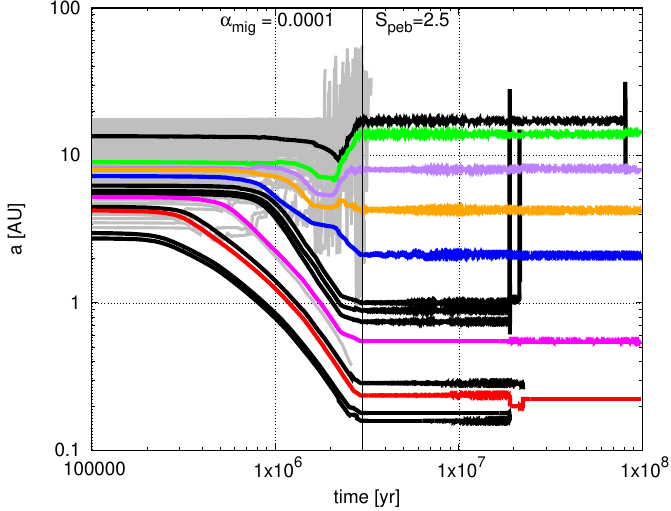}
 \includegraphics[scale=0.7]{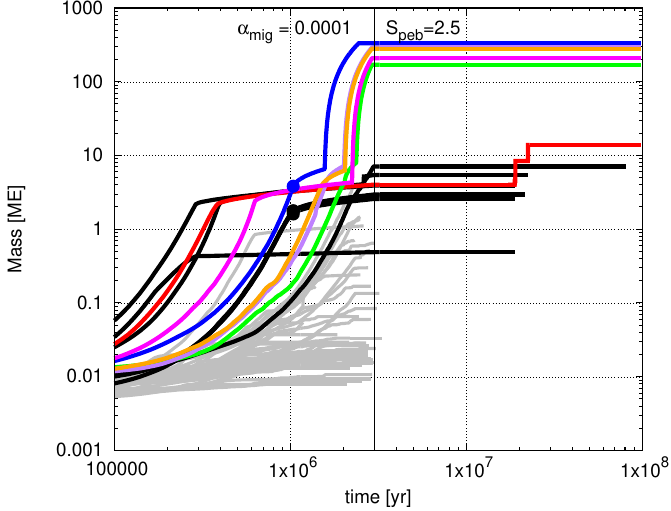}
 \includegraphics[scale=0.7]{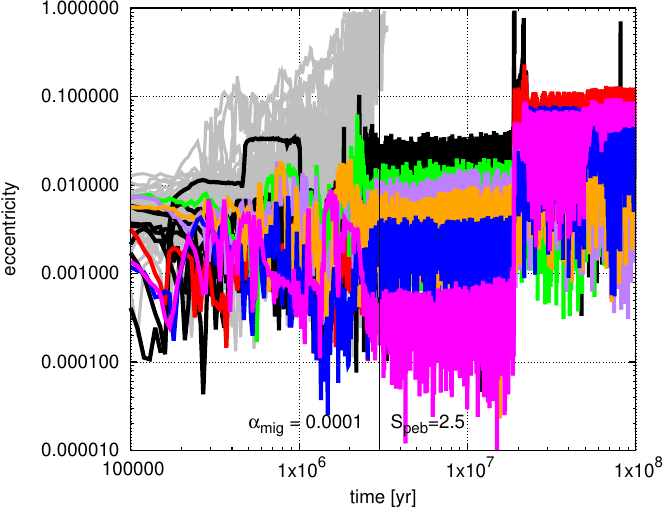}
 \includegraphics[scale=0.7]{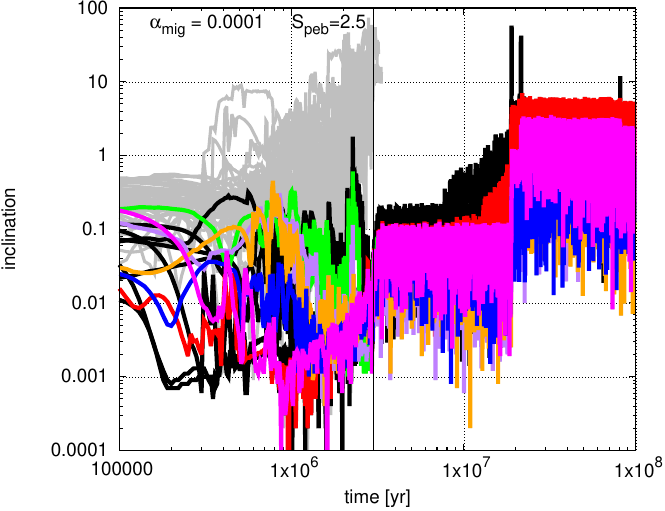}   
 \caption{Evolution of a system that underwent a late but relatively modest instability. Semi major axis (top left), planetary mass (top right), eccentricity (bottom left) and inclination (bottom right) of 60 planetary embryos as function of time. The gas disc lifetime is 3 Myr after injection of the planetary embryos, which migrate with the \citet{2018arXiv180511101K} prescription with $\alpha_{\rm mig}=0.0001$ and grow with a $S_{\rm peb} = 2.5$. The blue dot in the planetary mass plot marks when the planet indicated by the blue line reaches pebble isolation mass, blocking the flux of pebbles to the inner disc. At this time, the black bodies stop accreting pebbles as well (black dot) and thus start to slow contract their envelope. The black, grey and coloured lines show the same behaviour as bodies in Fig.~\ref{fig:largedistnominalmig}. At the end of the system evolution, the super-Earth (red) has an eccentricity of $\sim$0.1 and an inclination of a few degrees, while the remaining gas giants have eccentricities up to $\sim$0.05 and inclinations of only up to $\sim$1 degree.
   \label{fig:noedgenoscatter}
   }
\end{figure*}

\begin{figure*}
 \centering
 \includegraphics[scale=0.7]{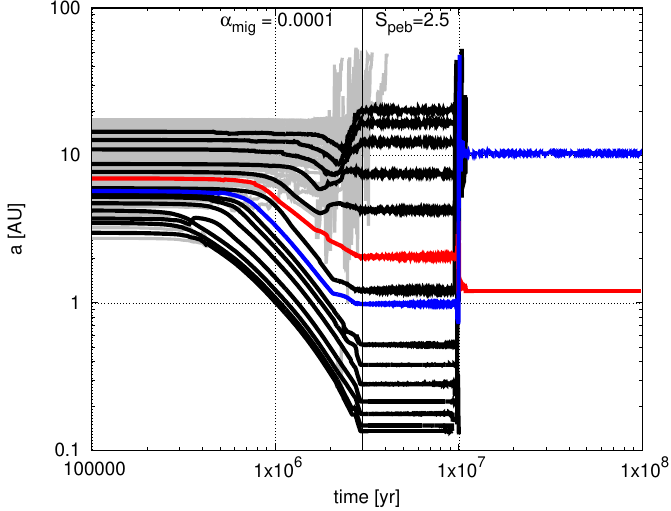}
 \includegraphics[scale=0.7]{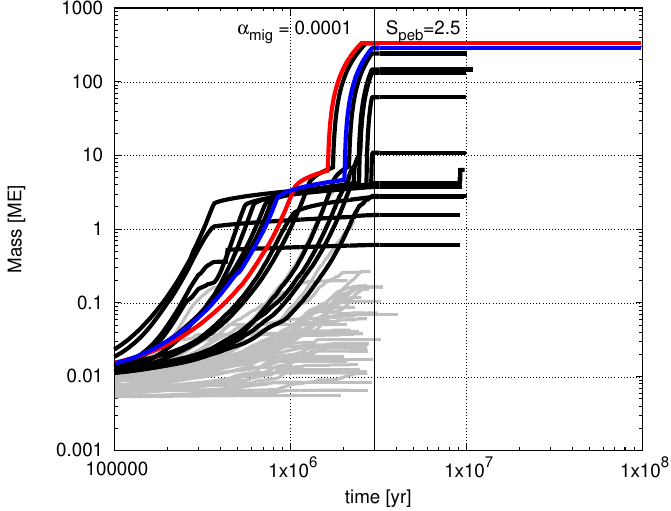}
 \includegraphics[scale=0.7]{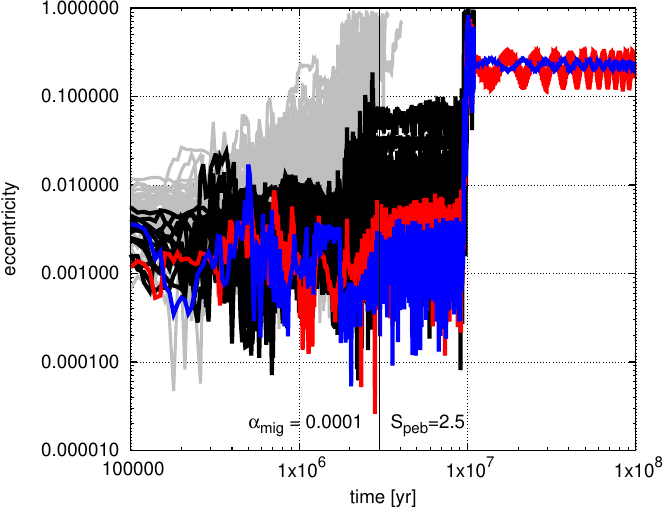}
 \includegraphics[scale=0.7]{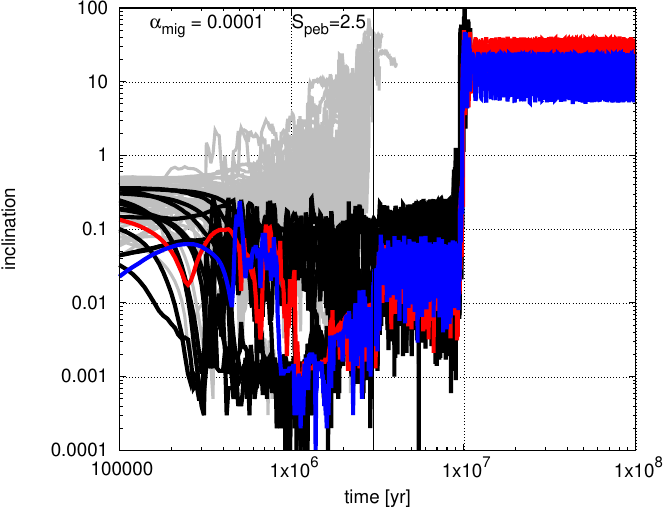}   
 \caption{Evolution of a system that underwent a late and dramatic instability. Semi major axis (top left), planetary mass (top right), eccentricity (bottom left) and inclination (bottom right) of 60 planetary embryos as function of time. The gas disc lifetime is 3 Myr after injection of the planetary embryos, which migrate with the \citet{2018arXiv180511101K} prescription with $\alpha_{\rm mig}=0.0001$ and grow with a $S_{\rm peb} = 2.5$. The black, grey and coloured lines indicate the same behaviour as bodies in Fig.~\ref{fig:largedistnominalmig}.
   \label{fig:noedgescatter}
   }
\end{figure*}

In Fig.~\ref{fig:noedgenoscatter} we show a system that is stable for a few 10 Myr, but then becomes unstable, where one super-Earth mass planets survives next to a few gas giants, while in Fig.~\ref{fig:noedgescatter} we show a system that undergoes a major instability, where only two gas giants survive on eccentric orbits. The simulations for the here presented systems use $S_{\rm peb}=2.5$ and $\alpha_{\rm mig}=0.0001$.

In Fig.~\ref{fig:noedgenoscatter} and Fig.~\ref{fig:noedgescatter} we show the evolution of the two systems. As some of the planets grow, the smaller bodies are excited in eccentricity and inclination. However, as soon as the gas disc dissipates at 3 Myr, the small bodies are ejected from the system and only bodies with at least $\approx$1 Earth mass survive.

As already discussed above, the planetary embryos originating from up to 5 AU start to grow first, but then also start to migrate first towards the inner disc. As they grow in a region of the disc, where the pebble isolation mass is small, they reach the pebble isolation mass very quickly (after a few 100kyr). As their pebble isolation mass is very small, they do not accrete gas efficiently and form a configuration of super-Earth mass planets in the inner planetary system. Planets growing in the outer system, have the potential to grow bigger. However, the outer planets form a pebble blockade. The planet indicated in blue in Fig.~\ref{fig:noedgenoscatter} reaches pebble isolation mass (marked by the blue dot) first and thus quenches growth by pebble accretion for the interior planets (black dot). These planets thus stay in the super-Earth mass regime and are too small to accrete a gaseous envelope during the gas disc phase. This behaviour is typical in our simulations and can be observed also in Figs.~\ref{fig:Kanagawamig00054} and \ref{fig:Kanagawamig0001}.

Exterior to the gas giant at 2 AU (blue curve in Fig.~\ref{fig:noedgenoscatter}), a group of three gas giants resides with orbits up to 15 AU. Even further away an ice giant of nearly 10 Earth masses has formed. The whole system configuration after gas disc dissipation at 3 Myr has some similarities with the WASP-47 system \citep{2015ApJ...812L..18B, 2016A&A...586A..93N}, where an interior super-Earth is orbited by a Jupiter type planet followed by another super-Earth and another Jupiter planet. Admittedly, the WASP-47 system is much tighter packed than the system in Fig.~\ref{fig:noedgenoscatter} after gas disc dissipation, but our simulations could explain a pathway to form a system like WASP-47. The evolution of the system in Fig.~\ref{fig:noedgenoscatter} is actually very similar to the evolution of the system shown in Fig.~\ref{fig:noedgescatter} until gas disc dissipation at 3 Myr.

\subsection{Long term evolution}

We focus here on the long-term evolution of the planetary systems and integrate the systems for 100 Myr to catch late instabilities in the systems. We study here only the simulations with $S_{\rm peb} = 2.5$ in order to investigate if super-Earths can survive in the systems with multiple giant planets.

The system in Fig.~\ref{fig:noedgenoscatter} contains 13 planets after gas disc dissipation within 20 AU, where 5 of those are gas giants. After about 20 Myr, the system undergoes a phase of instability, where as a consequence all the inner super-Earth planets except one are collided or are ejected from the planetary system. Additionally, after about 90 Myr, the ice giant is ejected as well. After 100 Myr of evolution only 1 super-Earth and the 5 gas giants survive. During these instabilities the eccentricities and inclinations of the planets are excited to moderate levels with eccentricities in the range of a few percent and inclinations up to a few degrees.

The system in Fig.~\ref{fig:noedgescatter} features 8 gas giants at the end of the gas disc's lifetime. This results in a major instability of the system a few Myr after gas disc dissipation. Towards 9 Myr, the inner super-Earths collide and form a body with an eccentricity of $\sim$0.8 and an semi-major axis of $\sim$0.15AU, which results in a perihelion very close to the central star\footnote{Tidal interactions with the central star can play a role here now, but are not included in our simulations. Nevertheless, we think their effects are minor in this case due to the short time until the instability in the outer system happens.} (where we assume a stellar radius of 0.01AU). The instability in the outer system at around 10 Myr then removes the outer gas giants and pushes the super-Earth in the central star. In the end only two gas giants survive.

The surviving gas giants orbit at 1.0 and 10 AU with eccentricities of 0.2-0.3. The final inclinations are around a few degrees up to even more than 10 degrees. The resulting system configuration with two gas giants, one close to 1 AU and one in the outer system resembles the structure of HD169830 \citep{2004A&A...415..391M} and HD183263 \citep{2009ApJ...693.1084W}. In both observed systems two gas giant planets recide on similar orbital configurations, including eccentricities exceeding 0.3. This type of evolution is comparable to previous simulations of planet-planet scattering \citep{2002Icar..156..570M, 2008ApJ...686..580C, 2008ApJ...686..603J, 2017A&A...598A..70S}.

\begin{figure}
 \centering
 \includegraphics[scale=0.7]{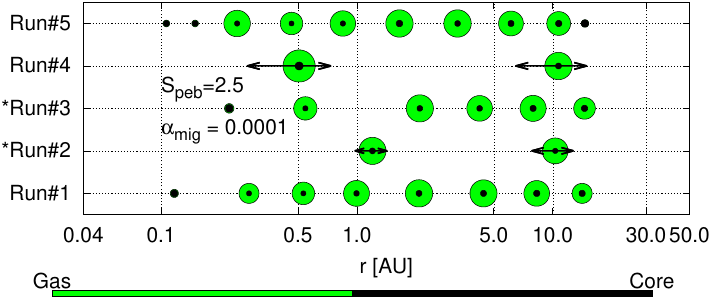}
 \caption{Final configurations after 100 Myr of integration of all our simulations with $\alpha_{\rm mig}=0.0001$ and $S_{\rm peb}=2.5$. The size of the circle is proportional to the total planetary mass (green) by the 3rd root and to the mass of the planetary core (black) also by the 3rd root. The runs marked with * are the example runs shown in Fig.~\ref{fig:noedgenoscatter} and Fig.~\ref{fig:noedgescatter}, while the black arrows indicate the aphelion and perihelion positions of the planet calculated through $r_{\rm P} \pm e\times r_{\rm P}$. The here presented systems are a long-term integration of the same systems shown after 8 Myr ($S_{\rm peb}=2.5$) of Fig.~\ref{fig:noedge0001}.
   \label{fig:longterm}
   }
\end{figure}

In Fig.~\ref{fig:longterm} we show the final configuration of our 5 simulations that we evolved for 100 Myr with $S_{\rm peb}=2.5$. The black arrows indicate the aphelion and perihelion distances of the eccentric planets. We only show them for the systems that underwent major instabilities, because all other eccentricities of the giant planets are well below 0.1. In our long term simulations 2 out of 5 systems undergo major instabilities, which is less than the fraction needed to explain the eccentricity distribution of giant planets, where $\sim$90\% of all systems need to undergo instabilities \citep{2008ApJ...686..603J}. This difference could be related to our still simplistic gas accretion routine or originate from a too slow/fast migration rate and will be investigated in future work.

There are two distinct outcomes, as described above. Either the systems undergo no instability (run 5) or only a slight instability (run 1 and 3), where some super-Earth planets are ejected from the system, but the overall system structure (inner super-Earth planet with outer gas giants) remains intact. In these systems, the giant planets are close to some higher order mean motion resonances (e.g. 2:7). New studies have actually suggested that the RV signal shown by single eccentric planets could have been mimiced by pairs of planets in resonance and current estimates suggest that up to 25\% of these eccentric Jupiter planets are in fact pairs of giant planets in mean motion resonances \citep{2018MNRAS.480.2846B}. Alternatively, the systems undergo major instabilities which result in the ejection of most planets (run 2 and 4) and only gas giants on very eccentric orbits remain.

We note that the systems that could resemble run 1 and run 5 are not observed at this point. The difference to the observations could be related to the planetary seed distribution and masses as well as to the gas contraction rates. We discuss these caveats in the following section. Alternatively, it could be that the timescale for instability is longer than the integration period of 100 Myr.

However, our simulations show that the formation of different planetary systems structures with gas giants is possible from the same initial disc structure and pebble fluxes. This result has not been found in previous planet population synthesis studies including pebble accretion, due to their limitation to one-planet per star \citep{2017ASSL..445..339B, 2018MNRAS.474..886N, 2018arXiv180810707B, 2018ApJ...864...77I, 2018arXiv181100523J}. The multi-body planet population synthesis simulations by \citet{2018ApJ...865...30C} show that indeed systems with small interior planets and outer gas giants can form in the same disc, similar to the here presented simulations.

\section{Discussion}
\label{sec:disc}

In this section we discuss caveats and implications of our giant planet formation scenario in respect to the exoplanet population and the solar system formation.

\subsection{Initial distribution of planetary embryos}

In our main simulations we initialise the planetary embryos starting from 2.75 AU outwards to 17.5 AU with an equal spacing of 0.25 AU between each planetary embryo. Additionally, the planetary embryos in our simulations are all within the same mass range. Our resulting simulations show that as soon as the pebble flux is high enough to allow the formation of gas giants, multiple gas giants (on average 3-4) form. This implies that our model overpredicts the number of gas giants when compared to the solar system, implying either that our understanding of the formation of ice giants \citep{2017ApJ...848...95V} and gas accretion \citep{2017MNRAS.471.4662C, 2017A&A...606A.146L} is incomplete  or that the solar system is an outlier, as suggested by observations. On the other hand, the general distribution of eccentricities of giant planets suggests that indeed when giant planets form at least 3 planets form to trigger large enough instabilities \citep{2008ApJ...686..603J}.

Additionally, many planets of at least a few Earths mass still migrate into the inner parts of the system, for all migration prescriptions attempted. These effects could have two reasons related to the initial distribution of planetary embryos:

\begin{enumerate}
 \item The starting mass of our planetary embryos is already in the regime of 0.01 Earth masses. However, when planetesimals form from the gravitational collapse of pebble clouds via the streaming instability, their size is roughly 100 km \citep{Johansen2015, 2016ApJ...822...55S, 2018ApJ...861...47S}. However, at this size, pebble accretion is not very efficient \citep{2016A&A...586A..66V, Johansen2017}. This implies that an initial stage of planetesimal collisions is needed to bring the planetary embryos to masses where pebble accretion can take over \citep{Johansen2017}. However, the collision rates between planetesimals scale with orbital distance, meaning that growth by planetesimals is less efficient in the outer disc, indicating that planetary embryos could be smaller in the outer regions of the disc. Less massive embryos in the outer disc result in longer growth timescales for these bodies, which might prevent them to grow all the way to gas giants and might leave them stranded as ice giants or even smaller bodies instead. A different initial planetary mass distribution might thus influence the final shape of the planetary system and will be studied in future work\footnote{ Test simulations with planetary embryos of 0.001 Earth masses and $S_{\rm peb}=2.5$ have confirmed that planets in the outer disc grow slower compared to planetary embryos with 0.01 Earth masses. Additionally, for smaller starting masses, the planetary embryos with a slightly larger initial mass grow quicker and dominate the system, due to their enhanced growth rate compared to their competitors.}.
 \item At ice lines, condensation can help to increase the size of pebbles and aid planetesimal formation \citep{2013A&A...552A.137R, 2017A&A...602A..21S}. Additionally, \citet{2016A&A...596L...3I} showed that rapid planetesimal formation just inwards of the water ice line is possible due to silicate-dust grains ejected from sublimating pebbles, eventually leading to the formation of dust-rich planetesimals directly by gravitational instability. Thus, the region around the water ice line could be the preferential location for the formation of planetesimals \citep{2016ApJ...828L...2A, 2017A&A...608A..92D}. This results in a narrow distribution of planetesimals and planetary embryos around the respective ice line instead of an equal distribution of planetary embryos throughout the whole disc. In turn this could lead to the formation of less gas giants that migrate into the inner disc. If that was the case for the formation of the solar system, a mechanism that then moves the giant planets outwards to their current positions like in the Grand Tack scenario \citep{2011Natur.475..206W} needs to be invoked.
\end{enumerate}


In our disc model the water ice line is already at around 1 AU, given the low disc accretion rate. However, in discs with higher accretion rates, the water ice line can reside at $\approx$5-6AU \citep{2015A&A...575A..28B}. If massive seeds would then only form around the water ice line, but nearly no planetary embryos interior to it\footnote{Planetary embryos growing in the inner disc only by the accretion of silicate pebbles grow very slow and do not form super-Earth planets, if outer planets are present \citep{2015Icar..258..418M, Izidoro18}.}, the growing planetary embryos at the water ice line could form the gas giants of the solar system without larger bodies migrating into the inner system due to the lack of growing bodies in this region. For this to work, higher initial disc accretion rates than used in the here presented simulations are needed to have a self-consistent water ice line position and planetary embryo starting position. Additionally, a steep radial decrease of the initial masses of the embryos is required to slow down the growth of outer bodies, which would then only grow to planets of a few Earth masses that could then collide and form the ice giants in the solar system \citep{2015A&A...582A..99I}. 

Recent analysis of observational data of systems hosting super-Earths and gas giants suggest that most systems with cold Jupiter should also host interior super-Earths, especially if the metallicity is above solar \citep{2018arXiv180502660Z, 2018arXiv180608799B}, but also that the opposite could be true \citep{2018arXiv180408329B}. Our simulations and those of our companion paper by \citet{Izidoro18} indicate that both outcomes are possible (see Fig.~\ref{fig:longterm}), thus casting doubt on the conclusions of \citet{2018arXiv180502660Z} that large-scale and substantial migration of solid material across discs is not possible.

However, our models additionally suggest that giant planets can only form if the pebble isolation mass is large (appendix~\ref{ap:Miso}), so in the outer disc. Systems without giant planets should thus have mostly embryos forming only in the inner regions of the disc or should feature low pebble fluxes. The initial distribution and masses of planetary embryos thus is key to understand the formation pathway of different planetary system. We will investigate how the distribution and initial masses of planetary embryos influences the formation of planetary systems in future work.

\subsection{Planet migration}

Our N-body simulations with nominal type-II migration confirm previous simulations of single bodies that gas giants that stay outside of 1 AU at the end of the gas disc's lifetime must have originated from far distances of 20-40 AU. Giant planets forming interior to this distance migrate to the inner edge of the disc forming hot Jupiters using the migration rates of our nominal model (Fig.~\ref{fig:largedistnominalmig}). During this process the formed Jupiters would destroy inner planetary systems \citep{2005A&A...441..791F, 2006Sci...313.1413R, 2007ApJ...660..823M}. Additionally these simulations would predict that most observed systems should host hot Jupiters. This is at odds with observation of exoplanetary systems. However, this result is based on the assumption that planetary seeds form all over the disc and with similar masses (see also above).

How much a planet migrates through a disc is related to the lifetime of the protoplanetary disc as well as to the starting time of the planetary seed. In our simulations, we use disc lifetimes of 3 Myr, which correspond to average disc lifetimes inferred by observations \citep{2009AIPC.1158....3M}. Shorter disc lifetimes would result in less time that the planets migrate, while longer disc lifetimes result in longer times that the planet can migrate.

Recent simulations of disc structure and evolution including magnetic fields, on the other hand, show that the accretion flow of the disc are mainly transported through the surface \citep{2016ApJ...821...80B, 2016arXiv160900437S}. In these scenarios, the mid-plane regions of the disc would be laminar and feature a low viscosity. However, studies of what the exact type-I and type-II migration rates in those discs are still under investigation and depend on the magnetic field strength \citep{2017MNRAS.472.1565M, 2018MNRAS.477.4596M}. We thus assume low viscosities for migration, testing two different configurations with $\alpha_{\rm mig}=0.00054$ and $\alpha_{\rm mig}=0.0001$, while keeping all disc parameters constant with $\alpha_{\rm disc}=0.0054$. We want to emphasise here again, that a lower $\alpha_{\rm mig}$ parameter not only reduces the type-II migration rate, but also allows an earlier transition into it. This is necessary to keep planets exterior to 1 AU (see appendix~\ref{ap:migration}).

We find in our simulations a clear trend between the migration speed and the location where planetary embryos need to originate to form gas giants with final semi-major axis larger than 1 AU. For $\alpha_{\rm mig}=0.00054$, the innermost origin of planetary embryos that form gas giants exterior to 1 AU is at $\approx 10$AU (Fig.~\ref{fig:Kanagawamig00054}), while for $\alpha_{\rm mig}=0.0001$ the origin of gas giants that stay exterior to 1 AU is at $\approx 5$AU (Fig.~\ref{fig:Kanagawamig0001}).

Reducing the migration speed even further would not move the origin of gas giants closer to the host star. This is caused by the very small pebble isolation mass in the inner disc (Fig.~\ref{fig:HrMiso}), which is so small that planetary cores do not contract their envelope easily (see also below). This indicates that if planetary embryos form all over the disc, low migration speeds are required to avoid an inward migration of gas giants from beyond 5 AU to orbits interior of 1 AU. Alternatively, scenarios like the Grand Tack scenario \citep{2011Natur.475..206W, 2014ApJ...795L..11P}, where the giant planets migrate outwards in resonance, are needed to avoid the inward migration of massive planets.

\subsection{Dependency of planet growth on the pebble flux}

The companion paper by \citet{Lambrechts18} identified a difference in the pebble flux by a factor of $\approx$1.7 that unveils a change in the growth mode of super-Earth planets that then either form by collisions from Mars size embryos (low pebble flux case) or directly by accreting pebbles efficiently during the gas-disc phase (high pebble flux scenario). As already stated in the companion paper by \citet{Izidoro18}, we find that a change of the pebble flux by a factor of $\approx 2-3$ can be enough to trigger the formation of gas giants exterior to 5 AU. 

In fact, the pebble flux proposed in \citet{Lambrechts18} to switch to the super-Earth growth mode is similar to the pebble flux needed in the here presented simulations to trigger gas giant formation (about a total of $\sim$170 Earth masses of pebbles). This difference in outcome is related to the Stokes number of the particles, which is larger by an order of magnitude in the here presented work. The reason why the Stokes number is larger in the here presented work compared to \citet{Lambrechts18} is related to the fact that we study here planet formation in the cold part of the disc, where water ice increases the sizes of the pebbles. In contrast, \citet{Lambrechts18} studies the formation of rocky super-Earths in the inner disc where pebbles are presumably chondrule size (mm). This difference of the Stokes number then translates directly into large differences in the accretion rate $\dot{M}_{\rm core}$. The requirement of \citet{Lambrechts18} to have a total of 170 Earth masses in silicate pebbles corresponds to 350 Earth masses in pebbles ($S_{\rm peb}=$5.0) in our simulations, due to the volatile loss at the water ice line.

A further increase of the pebble flux results in a slight increase in the efficiency of giant planet formation, as predicted by the metallicity correlation of giant planet observations \citep{2005ApJ...622.1102F, J2010}. This correlation was already shown in pebble accretion simulations with single planetary embryos \citep{2018MNRAS.474..886N}. At the same time, even in simulations with larger pebble flux, the formation of super-Earths in the inner system is possible, because the pebble isolation mass is very small in that region (see appendix~\ref{ap:Miso}). In this scenario, the inward migrating super-Earths pile up in resonance chains that can become unstable due to dynamical interactions. The remaining planetary systems match the observation of super-Earth systems, which is discussed in our companion paper by \citet{Izidoro18}.

This thus implies that not only the pebble flux, but also the birth environment of the embryo determines if a super-Earth or a giant planet emerges.

\subsection{Disc lifetimes and embryo formation time}

\citet{2017LPI....48.1386K} concluded through meteoritic evidence that the reservoirs of non-carbonaceous meteorites and carbonaceous meteorites were spatially separated in the protoplanetary disc around the young sun at about $\approx$1 Myr. This separation can be achieved by a growing planet that stops the inward flux of particles, which would correspond to Jupiter's core exceeding the pebble isolation mass (although the gas flow through the gap could carry along a significant amount of small dust (e.g. \citealt{2018arXiv180102341B, 2018ApJ...854..153W}) and mix the inner and outer reservoirs, violating the \citet{2017LPI....48.1386K} picture of two distinct reservoirs). Additionally, the recent comparison of observed mm-dust masses in protoplanetary discs with masses of observed planets lead to the conclusion that discs do apparently not show enough material in pebbles to explain the observed planet population \citep{2018arXiv180907374M}. This and the results by \citet{2017LPI....48.1386K} could hint that planet formation happens early, possibly within the first Myr of the disc lifetime. 

Our used disc model uses an initial accretion rate of $5\times10^{-9}{\rm M}_\odot$/yr, corresponding to an age of 2 Myr in \citet{2015A&A...575A..28B} and $1\times10^{-9}{\rm M}_\odot$/yr at 5 Myr, which is when the disc dissipates. These accretion rates are on the lower end of the observed disc accretion rates \citep{2016A&A...585A.136M}. Additionally they feature a water ice line position around $\approx$1AU, which is too close to the central star to form the Earth dry. In fact, \citet{2016Icar..267..368M} suggested that Jupiter blocks the icy pebbles early, when the disc's accretion rate is higher than in our here used model and the ice line is far out. This leaves the Earth dry even as the water ice line evolves.

Test simulations (not displayed) with higher initial disc accretion rates (a few $10^{-8}{\rm M}_\odot$/yr) have revealed the same trend as observed in the here presented simulations:
\begin{itemize}
 \item Higher disc accretion rates lead to faster inward migration for the same $\alpha$ parameter, due to the linear scaling of type-I migration with the gas surface density. In systems with higher accretion rates, planets migrate inwards faster and our simulations show more planets migrating interior of 1 AU. Reducing $\alpha_{\rm mig}$ further results in less migration of gap opening planets and similar system configurations as shown here can be produced.
 \item If discs then still evolve down to accretion rates of $1\times10^{-9}{\rm M}_\odot$/yr in 5 Myr, the planets have longer time to migrate, leading to more planets invading the inner system. Using shorter disc lifetimes ($\sim$3Myr) and higher initial disc accretion rates, we recover our results shown here, namely that planetary embryos starting at 7-9AU in discs with $\alpha_{\rm mig}=0.0001$ can stay exterior to 1 AU.
 \item Discs with higher gas accretion rates also feature correspondingly larger pebble fluxes in our model. This means the parameter $S_{\rm peb}$ does not need to be increased to values larger than $>2.5$ to allow the formation of gas giants in the outer disc, if the initial accretion rates are high. Using the nominal pebble flux of discs with higher accretion rates over 3 Myr leads again to a total pebble mass of 170 Earth masses, corresponding to $S_{\rm peb}=2.5$ used here in discs with lower accretion rates.
\end{itemize}

\subsection{Pebble accretion efficiency}

When a planet grows in a protoplanetary disc via pebble accretion, the planet only accretes a fraction $f_{\rm acc}$ of the pebble flux passing it (eq.~\ref{eq:facc}). Typically giant planets formed in our simulations have core masses of 5-15 Earth masses. Defining the core-mass-to-total-pebble-mass fraction, we can estimate how much of the total amount of pebbles is converted into planets
\begin{equation}
 f_{\rm p2p} = \frac{\sum_{\rm i}^N M_{\rm core,i}}{M_{\rm peb,tot}} \ .
\end{equation}
Here the $\sum_{\rm i}^N$ denotes the sum over all $N$-bodies present in the simulations. For single planet simulations $f_{\rm p2p}$ is thus very low, while it can increase up to $30\%$ for multi-planet systems. We note that we include in $f_{\rm p2p}$ also the solid material trapped in bodies that are later on ejected from the planetary system. We remind the reader that the total amount of pebbles drifting through the disc is 70 Earth masses for $S_{\rm peb}=1.0$.

We state the $f_{\rm p2p}$ ratio in Fig.~\ref{fig:noedge0001} for systems grown with a different total background pebble flux. In the simulations with $S_{\rm peb} = 1.0$ and $2.5$ about 25-30\% of the pebbles are accreted by the planets during the gas-disc lifetime. 

However, this percentage decreases with increasing $S_{\rm peb}$. This is caused by the higher total amount of pebbles available, but at the same time similar number of planets form with similar core masses. The planetary core mass is determined by the pebble isolation mass (appedix~\ref{ap:Miso}), and thus even at higher fluxes, the planetary core can not grow beyond this mass. At the same time, a similar number of planets grow in the disc at all pebble fluxes. Even though 60 planetary embryos are available in each simulation only 5-15 planetary embryos grow to fully grown planets, because the other embryos are scattered or ejected during the growth phase, as already noticed in the simulations without planet migration by \citet{2015Natur.524..322L}. The number of ejected planetary embryos is roughly the same in all simulations, independently of the pebble flux. The total mass of solids bound in planetary cores is then very similar in all simulations, but as the pebble flux increases $f_{\rm p2p}$ decreases.

We note that the $f_{\rm p2p}$-ratio just includes the amount of pebbles converted into planets in our simulations. In reality, pebbles can pile-up outside of planets that have reached pebble isolation mass \citep{2012A&A...546A..18M, 2014A&A...572A..35L} and form new planetesimals, which is not taken into account in our simulations. Our calculation thus reflects a minimum pebble-to-planet conversion ratio. Additionally, the pebble-to-planet conversion rate increases for simulations with lower pebble scale height, because giant planet formation is already possible at lower pebble fluxes (appendix~\ref{ap:Hpeb}).

This implies that pebble accretion is not very efficient to grow one single planetary core, but it is {\it quite efficient} for the formation of multiple planetary cores and thus for systems with multiple bodies \citep{2014A&A...572A.107L}. The problem of the low pebble accretion efficiency introduced in \citet{2018MNRAS.480.4338L} is therefore, in our opinion, simply a consequence of the consideration of single-core growth in \citet{2018MNRAS.480.4338L}.

\subsection{Envelope contraction}

The planets formed in the inner few AU in our simulations do not grow to gas giants, because their core masses are too small to trigger runaway gas accretion during the gas disc's lifetime (see appendix~\ref{ap:Miso}). This is caused by the slow gas contraction rate of the envelope (eq.~\ref{eq:Mdotenv}), which depends crucially on the planetary core mass ($\dot{M}_{\rm env} \propto M_{\rm core}^{11/3}$). This implies that a slightly larger core mass could allow an efficient contraction of the envelope and thus a transition into runaway gas accretion. This is the reason why the planetary seeds that grow to become gas giants have to form exterior to 4-5 AU in our model.

Our model of envelope contraction is based on 1D simulations, where just recently 2D and 3D simulations of embedded small mass planets have revealed further detail of this process. \citet{2015MNRAS.446.1026O} studied the flow of gas of embedded low mass planets in isothermal discs and found that the gas flow structures from the disc penetrates deep into the planetary hill sphere and recycle the gas around it preventing the contraction of the envelope. However, relaxing the isothermal assumption, \citet{2017MNRAS.471.4662C} and \citet{2017A&A...606A.146L} found that the planetary envelope is bound around low mass planets. Nevertheless, the simulations by \citet{2017A&A...606A.146L} revealed that planetary cores need to be more massive than $\approx$10 Earth masses before contraction becomes efficient. 

Collisions between super-Earths bodies can also increase the planetary masses to above pebble isolation masses, which then reduces the envelope contraction time. However, in our simulations collisions between super-Earth type planets at the inner disc edge at $\sim 0.1$AU mostly happen after gas disc dissipation, similar as in \citet{Izidoro18}.

On the other hand, our gas contraction rates allow planets that have reached just 5 Earth masses to contract their envelope and grow into a gas giant within 1 Myr (Fig.~\ref{fig:noedgenoscatter}). This indicates that our gas accretion rate may in fact be too efficient, and most of the planets that migrated into the inner system should stay icy super Earths as outlined by \citet{Izidoro18} and not grow into gas giants. The results presented in the paper by \citet{Izidoro18} thus remain valid. The here presented results, taking a less generous gas accretion recipe into account would imply that a pebble flux larger than $S_{\rm peb}=2.5$ and disc structures with larger aspect ratios to increase the pebble isolation mass might be needed to form giant planets.

Using the same pebble and gas accretion recipe as in this work, \citet{2018MNRAS.474..886N} showed that our gas accretion scheme is too efficient and too many gas giants are formed compared to observations. This was also confirmed by \citet{2018arXiv180810707B}. Additionally, \citet{2018arXiv181001389O} proposed a new mechanism that radial mass accretion in a disc can limit the gas accretion onto super-Earth cores, so that super-Earths interior to 1 AU do not grow into gas giants.

The results of \citet{2017A&A...606A.146L} also crucially depend on the opacity in the planetary envelope, where a lower opacity promotes faster envelope contraction. Our planetary envelope contraction model depends inversely to the opacity in the envelope (eq.~\ref{eq:Mdotenv}), meaning that a lower opacity results in larger accretion rates. In our simulations we use a fixed envelope opacity in the planetary atmosphere motivated by previous studies \citep{2008Icar..194..368M}. In principle, a reduction of the envelope opacity can thus dramatically reduce the contraction phase and even allow bodies of just a few Earth masses to reach runaway gas accretion (see also the appendix of \citealt{2015A&A...582A.112B}). However, we deem this scenario unrealistic, because otherwise most of the super-Earth planets would turn into gas giants, even in the inner disc, whereas observations predict that 30-50\% of all systems host close-in super-Earths \citep{2013ApJ...766...81F}. Note however that rocky planets up to approximately 5 Earth masses can form from small Mars-sized embryos after disc dissipation in a fashion similar to the terrestrial planets \citep{Lambrechts18}. In summary, a change of the opacity in the planetary envelope could thus change our giant planet formation frequency and will be investigated in the future.

\subsection{Cold gas giant formation}

In the previous subsections we discussed the conditions needed for giant planet formation (high pebble flux, formation outside of 5 AU, fast enough envelope contraction rates), while we here want to discuss the frequency and implications of gas giant formation for planetary systems. In our model, if giant planets form, several giant planets form, with a slight increase of the giant planet formation rate with pebble flux.

Due to the inward migration during the gas disc phase, the planets are relatively tightly packed. At the end of the gas-disc phase, the smaller bodies that did not grow beyond 0.1-1.0 Earth masses are ejected due to the lack of gas damping. The remaining systems can harbour even more than 10 planets at the end of the gas disc phase (see Fig.~\ref{fig:noedgenoscatter}). However, after the gas disc dissipation, the systems can become unstable, where we observe two main outcomes in our simulations:
\begin{enumerate}
 \item The smaller planets are ejected, but the gas giants remain on relatively circular outer orbits (Fig.~\ref{fig:noedgenoscatter}).
 \item The whole system undergoes an instability and the remaining giant planets orbit their host star on very eccentric orbits (Fig.~\ref{fig:noedgescatter}).
\end{enumerate}
It seems that a difference in these scenarios is the number of gas giants that are present in the disc after gas disc dissipation, where more gas giants are present that undergo massive instabilities. A higher pebble flux results in a higher formation frequency of giant planets, which in turn leads to more instabilities so that the remaining gas giants are more likely on highly eccentric orbits. This is in agreement with recent radial velocity observations that showed that eccentric cold Jupiters are more likely to orbit stars with higher metallicity \citep{2018arXiv180206794B}.

Additionally, during the gas-disc phase, the damping of eccentricity and inclination influences the stability of the planetary systems. Long damping time-scales promote instabilities, while short damping time-scales promote stability during the gas-disc phase. Our short damping time-scales are motivated by the 3D hydrodynamical simulations of \citet{2013A&A...555A.124B}, which are about an order of magnitude shorter than then classically assumed K-damping time-scales
\begin{equation}
 \dot{e}/e = - K_{\rm damp} |\dot{a}/a| \ ,
\end{equation}
with $K_{\rm damp}=100$, where typically $K_{\rm damp}=1-100$ \citep{2002ApJ...567..596L}. Using these short damping time-scales, \citet{2017A&A...598A..70S} showed that migration and scattering events of giant planets during the gas phase can reproduce the eccentricity distribution of giant planets. Previous works, using different giant planet damping time-scales, have concluded similarly \citep{2008ApJ...686..603J, 2008ApJ...688.1361M, 2009ApJ...691.1764M}.

Nevertheless, before trying to match the giant planet population within the framework of our simulations including pebble accretion and planet migration, the exact migration and damping rates of multiple planets embedded in discs driven by disc winds need to be derived.

\subsection{Hot gas giant formation}

Recent observations of hot Jupiters have tried to constrain the heavy element content in these planets \citep{2016ApJ...831...64T}, giving constraints to planet formation theories. Our simulations with $S_{\rm peb}=2.5$ and $\alpha_{\rm mig}=0.0001$ have the potential to keep giant planets exterior to 1 AU, if their planetary seeds form exterior to 5 AU. However, these simulations do not produce any hot Jupiter planets. 

We do not observe any scattering events of gas giants into the inner disc where they could become hot Jupiters. This seems to be in contrast to observations \citep{2018arXiv180206794B}, where scattering was suggested to be the main formation pathway of hot Jupiters. Instabilities on the other hand could occur on longer time-scales than the here considered integration \citep{2012ApJ...751..119B}. The recent review by \citet{2018ARA&A..56..175D} additionally highlighted that all proposed formation chanels (in-situ formation, migration and scattering) have problems explain the properties of hot Jupiters, emphasising the complexity of the problem. We note here that our simulations at this point do not aim to reproduce the hot Jupiter population and are thus also missing important physics needed to explain hot Jupiter formation by scattering, for example tidal effects. Nevertheless we want to discuss the implications of the two formation chanels of hot Jupiters from our simulations:
\begin{enumerate}
 \item Large pebble fluxes ($S_{\rm peb}>5.0$), which allow fast growth of the planets in the inner disc to reach runaway gas accretion (Fig.~\ref{fig:noedge0001}). 
 \item Fast inward migration as illustrated in section~\ref{sec:nominal} allows the penetration of super Earths and gas giants into the inner system.
\end{enumerate}
However, both scenarios have some flaws. Scenario 1 does not always lead to the formation of hot Jupiters, even for large pebble fluxes. Additionally, the core masses and thus potentially the heavy element content (assuming that not too many heavy elements can be accreted through the gas phase, but see \citealt{2017MNRAS.469.3994B}) is very low in contradiction to \citet{2016ApJ...831...64T}.

In the second scenario where hot Jupiters can be formed by the rapid inward migration of super-Earths and gas giants allows for core masses in agreement with the heavy element content found by \citet{2016ApJ...831...64T} due to mutual collisions at the discs inner edge during the gas disc phase. This formation pathway is also possible at low pebble fluxes (see also \citet{Izidoro18}), representing low metallicity environments, where hot Jupiters are actually rare \citep{2018arXiv180206794B}.

On the other hand, if the second formation pathway with fast migration rates was the norm, then systems of cold gas giants should be very rare, in contradiction to observations.

\subsection{Solar system formation}

Our simulations normally form multiple planets of at least a few Earth masses and many planets migrate into the inner disc, inconsistent with the solar system formation (see above), but in agreement with exoplanet systems (see the companion paper by \citet{Izidoro18}). At the same time, the formation of ice giants is quite rare. This can be related, as discussed above, due to the initial masses of the planetary embryos. Additionally, the formation of ice giants seems only possible in the simulations, where the pebble flux is 2.5 times the nominal pebble flux. Larger pebble fluxes turn the ice giants into gas giants, because their cores can grow quicker, allowing more time for envelope contraction, see above. This implies that for the formation of the solar system a pebble flux that allows the formation of gas giants around 5 AU, but at the same time does not allow the growth of gas giants around 10 AU and limits the growth of those bodies to ice giants instead is required. In the pebble accretion picture, ice giants did not reach pebble isolation mass, but accreted a polluted envelope consisting mostly of hydrogen during the gas disc phase while accreting pebbles \citep{2014A&A...572A..35L, 2017ApJ...848...95V}. We note that even though the number of our simulations limited, the formation of outer ice giants is also quite rare in planet population synthesis models that include the same growth and disc model \citep{2017ASSL..445..339B, 2018MNRAS.474..886N}. This implies that the formation of ice giants in this framework might indeed be hard, but solutions (as discussed above) will be investigated in the future.

The Earth formed after the gas-disc phase \citep{2009GeCoA..73.5150K}. \citet{2015Icar..258..418M} suggested that Jupiter's core formed efficiently outside the ice line due to the accretion of water rich pebbles. Then Jupiter reached pebble isolation mass, blocking the inward flux of pebbles and quenching the growth of the inner embryos in the terrestrial planet zone. Our companion paper by \citet{Lambrechts18} shows that under such a reduced flux of pebbles in to the terrestrial zone, interior embryos that only grow to roughly 0.1 Earth masses can then form terrestrial planets through collisions of embryos after the gas-disc phase.



In the solar system, Jupiter and Saturn orbit at 5.2 and 9.7 AU, which is farther away from the central star than most gas giants formed in our simulations. However, from the formation perspective of the solar system, the gas giants could have migrated to the inner disc (Jupiter migrating inwards to 1.5-2.0 AU) and then migrated outwards in resonance \citep{2011Natur.475..206W, 2014ApJ...795L..11P}. This would solve the problem of inward migration of gas giants close to 1 AU in our model in the framework of the formation of the solar system. Just stalling the migration of the giant planets would still results in the planets penetrating into the inner systems, as they migrate inwards mostly during their type-I migration phase. However, these effects are not included in our models, but we will investigate these effects in the future. Without reducing the migration rates even further, the Grand Tack scenario seems necessary to explain the formation of the solar systems, if Jupiters core formed at $\sim$5 AU.

\section{Summary}
\label{sec:summary}

In this paper we present results of N-body simulations including pebble accretion and planetary migration in the framework of giant planet formation. Our study builds on the single body approach of \citet{2015A&A...582A.112B} regarding the formation recipes of giant planets. 

In the simulations with nominal migration rates, we confirm the result of \citet{2015A&A...582A.112B} that planetary embryos growing to form gas giants exterior to 1 AU have to originate from 20-40 AU using nominal planet migration rates. If planetary embryos form all over the disc, this scenario predicts a very large fraction of super-Earth to Jupiter-mass planets that migrate into the inner disc, in contrast to our own solar system (but see also section~\ref{sec:disc}).

In order to form giant planets exterior to 1 AU, our simulations indicate that certain conditions have to be met. In particular, the structure of the formed planetary systems show a strong dependency on the pebble flux, the migration speed and the pebble isolation mass.

The pebble accretion rate in itself depends on the surface density of pebbles at the planets location, which in turn scales with the square-root of the pebble flux (eq.~\ref{eq:SigmaPeb}). Thus, if the pebble flux is too low, the growth rates become too low and planets can not grow to reach pebble isolation mass within the gas disc's lifetime. We show that a certain pebble flux is needed to allow a fast enough accretion to form the cores of giant planets. In particular, the simulations with nominal migration rates show that a total mass of 700 Earth masses of pebbles is needed to form giant planets starting at 20-40 AU ($S_{\rm peb}=10.0$ in Fig.~\ref{fig:largedistnominalmig}).

The final position of growing planets is determined by migration. Recent studies of planet-disc interactions indicated that lower viscosities can allow an earlier transition to type-II migration and thus reduce large scale inward migration. We investigate how a lower viscosity for planet migration influences the formation of planetary systems, where we focus on two different migration speeds (Fig.~\ref{fig:Kanagawamig00054} and Fig.~\ref{fig:Kanagawamig0001}).

Our simulations indicate that a reduction to $\alpha_{\rm mig}=0.00054$ allows the formation of giant planets that stay exterior to 1 AU originating from 10 AU, while $\alpha_{\rm mig}=0.0001$ pushes this boundary down to 5 AU. However, reducing $\alpha_{\rm mig}$ to even lower values does not change this boundary, because the final mass of the planet is mostly independent of the pebble flux. Instead it depends on the disc's scale height, where the low scale heights in the inner disc cut growth via pebble accretion at a few Earth masses (see appendix~\ref{ap:Miso}). This prevents planetary embryos formed in this region to grow into gas giants and these planets stay in the mass regime of super-Earths. Additionally, this implies that the formation of super-Earths in the inner part of the disc is independent of the pebble flux as predicted by observations \citep{2012Natur.486..375B, 2015ApJ...808..187B}, as long as the minimal flux for the formation of super-Earths is reached (see \citealt{Lambrechts18}).

Additionally planetary embryos formed closer to the central star grow faster due to the reduced pebble scale height and larger pebble surface density. This implies that a smaller pebble flux is sufficient to form giant planets from embryos exterior to 5 AU, if the migration speed is low. Indeed, with lower migration rates a total pebble mass of less than 200 Earth masses is needed to allow embryos to grow to giant planets.

Our studies show that an increase in the total time-integrated pebble flux of up to $\sim$350 Earth masses in total allows the formation of more gas giants, in agreement with the observed metallicity correlation of giant planets \citep{2005ApJ...622.1102F, J2010}. However, for even larger pebble fluxes, the giant planet formation frequency does not increase any more, because of the efficient ejection of small bodies caused by the growing gas giants preventing growth of many more bodies.

In Fig.~\ref{fig:noedge0001} we show the final configuration of our planetary systems 5 Myr after gas disc dissipation as a function of time-integrated pebble flux using a migration speed set by $\alpha_{\rm mig}=0.0001$. Low pebble fluxes result in systems with only super-Earth mass planets, while systems with higher pebble fluxes allow the formation of gas giants. Intermediate pebble fluxes ($S_{\rm peb}=2.5$) allow the formation of systems that harbour inner super-Earths with outer gas giants.

We finally present results of full planetary systems, where we investigate also the long term evolution after gas disc dissipation. As our simulations normally show around 10 or more planets that formed during the gas disc phase, the resulting systems are quite tightly packed. After gas disc dissipation, the planetary eccentricity and inclination are not damped by the gas any more and the smaller bodies are ejected from the system. 

In Fig.~\ref{fig:noedgenoscatter} and Fig.~\ref{fig:noedgescatter} we present two typical outcomes of our simulations. Either the system undergoes a massive instability and only two Jupiter planets remain on very eccentric orbits, or the system undergoes a smaller instability in which process not only the small bodies, but also some formed super-Earth planets are ejected. The resulting systems could then still harbour several planets, which are then mostly on quite circular orbits.

The here presented results, as well as the results of our companion papers \citep{Lambrechts18, Izidoro18}, open new ways to study the formation of planetary systems by combining pebble accretion, planet migration and N-body dynamics. In the future we will expand our simulations to additionally study the evolution of full systems with terrestrial planets, super-Earths and gas giants. Future research should be directed to the open issues pointed out in section~\ref{sec:disc} and the work presented in this trilogy of papers can act as a reference.

\begin{acknowledgements}

B.B., thanks the European Research Council (ERC Starting Grant 757448-PAMDORA) for their financial support. A.J. was supported by the European Research Council under ERC Consolidator Grant agreement 724687-PLANETESYS, the Swedish Research Council (grant 2014-5775), and the Knut and Alice Wallenberg Foundation (grants 2012.0150, 2014.0017, and 2014.0048). S. N. R. and A.M. thank the Agence Nationale pour la Recherche for support via grant ANR-13-BS05-0003-002 (grant MOJO). A. I. thank FAPESP for support via grants 16/19556-7 and 16/12686-2. We thank the referee John Chambers for his report and comments that helped to improve this manuscript.

\end{acknowledgements}

\appendix
\section{Disc evolution and pebble isolation mass}
\label{ap:Miso}

The evolution of our disc model is described in detail in \citet{2015A&A...575A..28B}, where we follow the semi-analytical fit of the disc structure provided in the appendix of that work. For our planet formation simulations we chose a disc that is already evolved for 2 Myr according to the disc evolution prescription of \citet{2015A&A...575A..28B}. Therefore $t=0$ Myr in the here presented work corresponds to $t_{\rm B15} =2$ Myr.

\begin{figure}
 \centering
 \includegraphics[scale=0.7]{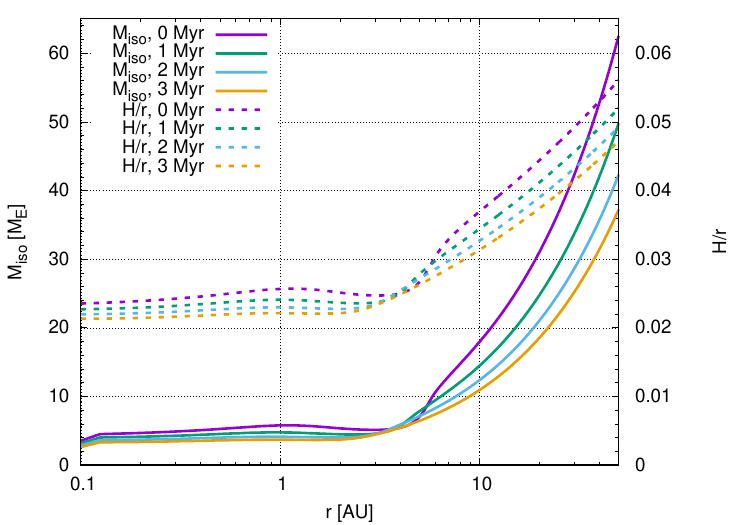}
 \caption{The pebble isolation mass $M_{\rm iso}$ (eq.~\ref{eq:MisowD}, solid lines) and the disc's aspect ratio $H/r$ (dashed) for different disc ages. As the disc cools in time, the pebble isolation mass is reduced. However, in the inner few AU of the disc, the pebble isolation mass stays constant at around 4-5 Earth masses, which is too low for efficient envelope contraction. Here $\alpha_{\rm disc}=0.0054$ has been used to calculate the pebble isolation mass. The age of 0 Myr corresponds to a disc that already evolved for 2 Myr in the \citet{2015A&A...575A..28B} disc model and corresponds to our starting disc age.
   \label{fig:HrMiso}
   }
\end{figure}

In fig.~\ref{fig:HrMiso} we show the discs aspect ratio and the resulting pebble isolation mass (eq.~\ref{eq:MisowD}) as a function of time. As the disc evolves in time, it cools down, however, at this stage the inner regions of the disc are not dominated by viscous heating any more, but by stellar irradiation, which only changes slightly during this evolution stage \citep{2015A&A...575A..28B}. Therefore the disc's aspect ratio is only changing slightly in time. In the inner few AU of the disc, the aspect ratio is nearly constant, while it rises with $H/r \propto r^{2/7}$ outside of 5 AU due to the efficient absorption of stellar irradiation.

As the pebble isolation mass $M_{\rm iso}$ depends cubicly on $H/r$, $M_{\rm iso}$ follows the disc structure, resulting in a roughly constant pebble isolation mass in the inner few AU and a significant increase exterior to 5 AU. In particular due to the low $H/r$, the pebble isolation mass is just between 4-5 Earth masses in the inner disc. However, these planetary masses are too low to start to efficiently accrete a gaseous envelope \citep{2017A&A...606A.146L}. Only exterior to 5 AU is the pebble isolation mass large enough for planets to start to accrete gas efficiently in our model, due to the strong dependency of the gas contraction rates on the planetary mass (eq.~\ref{eq:Mdotenv}). The contraction efficiency of planetary envelopes additionally depends on the opacity of the envelope and is an active area of research (see section~\ref{sec:disc}).

\section{Migration}
\label{ap:migration}

\begin{figure}
 \centering
 \includegraphics[scale=0.7]{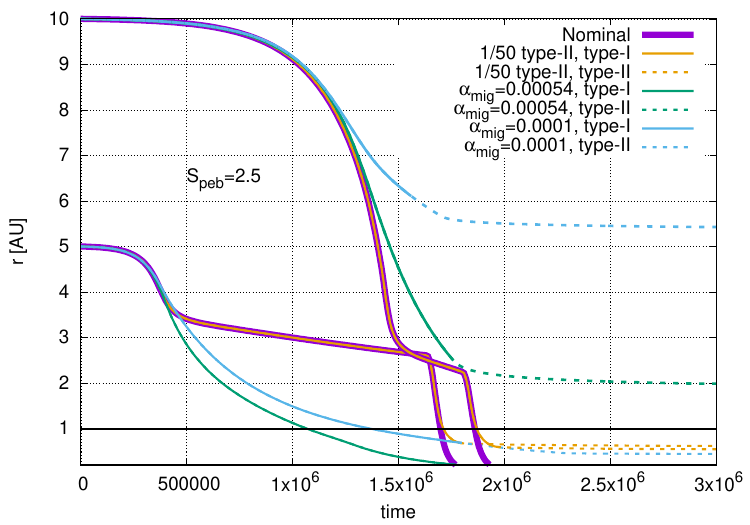}
 \caption{Semi major axis evolution of single planets starting at 5 or 10 AU in discs with $S_{\rm peb} = 2.5$. The bodies migrate at 4 different speeds, (i) the nominal migration speed (purple), (ii) with a nominal type-I migration but a type-II migration of 1/50 of the nominal type-II migration rate (yellow) and (iii) following the \citet{2018arXiv180511101K} prescription with $\alpha_{\rm mig}=0.00054$ (green) and $\alpha_{\rm mig}=0.0001$ (blue). Reduction of type-II migration does not change the migration path exterior to 1 AU compared to the nominal migration rates, because the transition to type-II migration is too late. The bodies that cross the 1 AU threshold have roughly masses of typical super-Earths to Neptune mass. We do not show the evolution of bodies crossing interior of 0.2 AU.
   \label{fig:semi1body}
   }
\end{figure}

In our work we investigate different migration prescriptions as outline above. To illustrate how these different migration prescriptions actually influence the movement of a body through the disc, we show in Fig.~\ref{fig:semi1body} the semi-major axis evolution of bodies starting at 5 and 10 AU. We use the nominal migration prescription used in \citet{2015A&A...582A.112B} and described in section~\ref{subsec:nominal}, a nominal type-I rate combined with a type-II migration rate that is 1/50 of the nominal type-II rate and the \citet{2018arXiv180511101K} prescription with $\alpha_{\rm mig}=0.00054$ and $\alpha_{\rm mig}=0.0001$ as described in section~\ref{subsec:kana}. The planets in Fig.~\ref{fig:semi1body} start at 0.01 Earth masses and grow through pebble accretion in a disc with $S_{\rm peb} = 2.5$ as they migrate. For the \citet{2018arXiv180511101K} prescription we determine type-II migration when the gap depth has reached 90\% of the unperturbed gas surface density value.

Using the nominal migration prescription without reduced viscosity allows planets to migrate outwards around 2-3 AU (purple line in Fig.~\ref{fig:semi1body}). But the planet in Fig.~\ref{fig:semi1body} start exterior to the region of outward migration, so these regions of outward migration will just slow down inward migration \citep{2015A&A...575A..28B}. As these regions of outward migration evolve, the planets follow their evolution. However, the region of outward migration can not contain the planet forever, so that it eventually grows too big and migrates inwards. For the planetary embryo staring at 10 AU, the region of outward migration delays the rapid inward migration as soon as the planet migrated inwards to around 3 AU. But, also then the planet becomes too large to be contained in the region of outward migration and migrates to the inner disc.

We now test the influence of a type-II migration rate reduced by a factor of 50 (yellow line). This reduction in type-II migration rate does not influence the growth tracks of planets exterior to 1 AU, because the planets do not open a deep gap and thus never transition into type-II migration exterior to 1 AU. Thus, in order to avoid large scale inward migration, not only does the type-II migration rate have to be lower, but also the transition into type-II migration needs to happen earlier during the planets evolution. 

In the following we thus use the \citet{2018arXiv180511101K} prescription with a reduced viscosity (implemented in our simulations through a reduced $\alpha_{\rm mig}$ parameter) to allow earlier gap opening and slower type-II migration (blue and green lines).

Using a reduced viscosity for migration, outward migration is not possible any more due to torque saturation. The planets migrating with the low $\alpha_{\rm mig}$ thus always migrate inwards. However, the lower $\alpha_{\rm mig}$ value allows an earlier gap opening and transition to type-II migration. This eventually saves the planets from migrating interior to 1 AU, if $\alpha_{\rm mig}$ is low enough and the planetary embryos start far away from the central star.

In Fig.~\ref{fig:semi1body} we see that planets forming at 5 AU are not saved from migrating interior to 0.2 AU for $\alpha=0.0001$. However, in Fig.~\ref{fig:Kanagawamig0001} planets forming at 5 AU with $\alpha_{\rm mig} = 0.0001$ survive in the disc outside of 1 AU. This is caused by the mutual interactions of the planets, reducing their growth (due to eccentricity excitement), but also the migration behaviour by trapping in resonances.

\section{Lower pebble scale height}
\label{ap:Hpeb}

In our simulations we calculated $H_{\rm peb}$ using $\alpha_{\rm disc}=0.0054$, even for the simulations using $\alpha_{\rm mig} \ll \alpha_{\rm disc}$, because we want to study the influence of migration on the formation of planetary systems alone. A reduced viscosity for pebble stirring reduces $H_{\rm peb}$ and will increase the planetary growth rates via pebble accretion. This is related to an earlier transition from 3D accretion to the faster 2D accretion due to the smaller pebble scale height \citep{2015Icar..258..418M}. In the following we thus test the influence on the formation of planetary systems if $H_{\rm peb}$ is calculated with $\alpha_{\rm mig}$.

\begin{figure}
 \centering
 \includegraphics[scale=0.7]{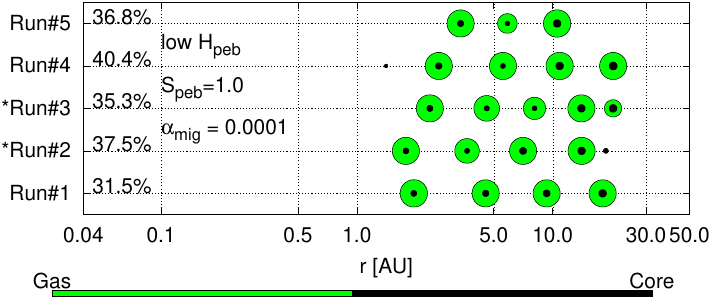}
 \caption{Final configuration of planetary systems after 8 Myr of integration using $\alpha_{\rm mig}=0.0001$ and a pebble scale height calculated with $\alpha=0.0001$. Planets migrating interior to 1 AU are removed from the simulations. We only show planets with masses larger than 0.3${\rm M}_{\rm E}$. The numbers in front show the pebble accretion efficiency of the whole system in percent.
   \label{fig:lowHpebcirc}
   }
\end{figure}

Using the nominal pebble flux rate we simulate the evolution of planetary systems using $\alpha_{\rm mig}=0.0001$ and display the final systems in Fig.~\ref{fig:lowHpebcirc}. As in section~\ref{sec:Kanagawamig} we remove planetary embryos migrating interior to 1 AU. The final systems are very similar to the planetary systems formed with $S_{\rm peb} = 2.5-5.0$ and larger $H_{\rm peb}$ displayed in Fig.~\ref{fig:Sim00054}. Indeed growth tracks of single planets are very similar for these type of simulations as well.

Additionally, as a similar number of planets are formed in the simulations with lower $H_{\rm peb}$ using the nominal pebble flux and in the simulations with higher $H_{\rm peb}$ using a higher pebble flux, the pebble-to-planet conversion rate is larger for the simulations with lower $H_{\rm peb}$. As indicated in Fig.~\ref{fig:lowHpebcirc} this conversion rate is now 30-40\%. This implies that pebble accretion becomes even more efficient in simulations with low vertical stirring.

Additionally, this implies that systems with multiple gas giants can form from only 70${\rm M}_{\rm E}$ of pebbles, which is less solid material than is incorporated into the giant planets in our solar system. On the other hand, the core masses in Fig.~\ref{fig:lowHpebcirc} are lower than for the giants in our solar system, again hinting that our gas accretion rates are probably too generous. At this point it is thus difficult to distinguish if giant planet systems form in discs with a high pebble flux ($S_{\rm peb}$) or with low $H_{\rm peb}$.

\bibliographystyle{aa}
\bibliography{Stellar}
\end{document}